%% file: elspaper.tex
\documentclass[preprint,12pt]{elsarticle}

\usepackage{kurz}
\usepackage{amssymb}
\usepackage{mathrsfs}
\usepackage{multirow}    
\usepackage{stmaryrd}
\usepackage{amsmath}
\usepackage{caption}  
\usepackage{a4wide}
\biboptions{numbers,sort&compress}
\usepackage{upgreek}
\usepackage{longtable}
\usepackage{xcolor}

\usepackage{arydshln}
\usepackage{mathtools}

\journal{International Journal for Numerical Methods in Engineering}

\begin{document}
\begin{frontmatter}



\title{Direct dissipation based arc-length approach for the cracking elements method}


\author[Hebei]{Yiming Zhang\corref{cor}}

\author[Hebei]{Junguang Huang}

\author[tj,tj2]{Yong Yuan}

\author[tuwien,tj,tj2]{Herbert Mang\corref{cor}}

\address[Hebei]{School of Civil and Transportation Engineering, Hebei University of Technology, Xiping Road 5340, 300401~Tianjin,~P.R.China }
\address[tj]{Department of Geotechnical Engineering, Tongji University, Siping Road 1239, 200092~Shanghai,~P.R.China}
\address[tj2]{State Key Laboratory for Disaster Reduction in Civil Engineering, Tongji University, Siping Road 1239, 200092~Shanghai,~P.R.China}
\address[tuwien]{Institute for Mechanics of Materials and Structures (IMWS), Vienna University of Technology,Karlsplatz 13/202, 1040 Vienna, Austria}
\cortext[cor]{Corresponding authors:\\\mbox{Herbert Mang}, \mbox{herbert.mang@tuwien.ac.at}\\
	                                 \mbox{Yiming Zhang}, \mbox{yiming.zhang@hebut.edu.cn}}

\begin{abstract}
\input{abstract}

\end{abstract}

\end{frontmatter}

\input{content}




\clearpage
\bibliographystyle{ieeetr}
\bibliography{Reference}







\end{document}

%% file: abstract.tex
Dissipated energy, representing a monotonically increasing state variable in nonlinear fracture mechanics, can be used as a restraint for tracing the dissipation instead of the elastic unloading path of the structure response.  In this work, in contrast to other energy-based approaches that use internal energy and the work done by the external loads, a novel arc-length approach is proposed.  It directly extracts the dissipated energy based on crack openings and tractions (displacement jumps and cohesive forces between two surfaces of one crack), taking advantage of the global/extended method of cracking elements.  Its linearized form is developed, and the stiffness factor of the arc-length restraint is naturally obtained by means of the Sherman-Morrison formula.  Once cohesive cracks appear, the proposed approach can be applied until most of the fracture energy is dissipated.  Results from several numerical tests, in which arc-length control and self-propagating cracks are jointly used, are presented.  They demonstrate the robustness of the proposed method, which captures both global and local peak loads and all snap-back parts of the force-displacement responses of loaded structures with multiple cracks.

%% file: content.tex
\section{Introduction}
Loading of structures, made of quasi-brittle material, results in a strong and almost instant stress release and redistribution in consequence of the strain softening process \cite{Nguyen1993}, starting when the local or global peak load is reached.  This process commonly results in numerical instability, characterized by snap-through and snap-back behavior at force or displacement control \cite{May2016}.  Path-following methods, among which the arc-length method is the most successful one, allow tracing the equilibrium path in a continuous and robust manner.

Arc-length methods commonly introduce a new unknown load control, which is treated together with the original balance equation.  Correspondingly, a new restraint equation is formed to obtain a monotonically increasing state variable during the loading process.  Therefore, the loading path can be smoothly followed, and failure of the analysis when tracing the elastic unloading path can be avoided.  The increment of the state variable, at each step, correlates to the the arc length.  Conventionally, the state variable is associated with the displacements.  When considering the global displacements, for example, adoption of the sum of the incremental displacements as the state variable , its monotonic increase cannot be assured.  It was shown that consideration of the local displacements, i.e., crack mouth displacements (opening or sliding), generally provides more stable results \cite{DEBORST1987211}.  Unfortunately, in engineering practice, the positions of the crack mouths are usually not known prior.  Therefore a more reliable state variable for automatically tracing the dissipation path is advantageous. 

For structures experiencing cracking, damage and dissipation of energy are irreversible processes.  Hence, the dissipation of energy is an ideal state variable for the arc-length method.  Early works presented in \cite{Gutierrez2004,Verhoosel2008} used the internal energy and work done by the external loads to obtain the dissipated energy.  Later, the authors of \cite{May2016} showed that the same procedure can be applied, with both the internal and dissipated energy following the entire loading path, where the switch of the arc-length state variable can be controlled by a scale factor.  Energy-based arc-length methods were successfully applied in  \cite{SINGH201614,OZDEMIR2019208,LABANDA2018319,Wang2015a}.  Moreover, a hybrid version was recently proposed in \cite{MEJIASANCHEZ2020405} as a combined displacement- and dissipation-based arc-length method to enhance the numerical stability.

The Cracking Elements Method (CEM) is a novel numerical approach for simulating quasi-brittle fracture \cite{Yiming:14,Yiming:16}.  Similar to the cracking particle method \cite{Rabczuk20102437,Rabczuk2004}, CEM uses disconnected cracking segments to represent cracking regions, which avoids precise descriptions of crack tips and prediction of crack paths (crack tracking) \cite{Saloustros2018}.  The CEM is a crack-opening based approach that uses the characteristic length like some of the damage-degree based models \cite{Yiming:19,Cervera:04}.  In a global treatment \cite{Yiming:20}, the crack openings of the CEM can be introduced as global degrees of freedom, making it possible to extract the dissipated energy directly.  

In this work, a direct dissipation-based arc-length approach for the CEM is proposed, where the dissipated energy is obtained on the basis of crack openings and tractions.  The stiffness factor of the arc-length restraint can be obtained by means of the Sherman-Morrison formula.  With this approach, the dissipation path can be followed from the appearance of the first crack until the collapse of the structure.  To assess the robustness of the approach, irregular meshes are considered and all cracks are self-propagating with the CEM.

The remaining parts of this paper are organized as follows: in Section~\ref{sec:cem}, the CEM is introduced briefly, and the traction-separation law, the kinematics and the numerical formulation are provided.  In Section~\ref{sec:arc}, the direct dissipation-based arc-length approach, including its basic restraints and the detailed numerical procedures, based on the Sherman-Morrison formula, is presented.  In Section~\ref{sec:ne}, numerical examples are given to demonstrate the robustness and the reliability of the approach.  Finally, Section~\ref{sec:conc} contains concluding remarks.

\section{The Cracking Elements Method}
\label{sec:cem}
The CEM was first presented in \cite{Yiming:14}.  It is based on the strong discontinuity embedded approach of the statically optimal symmetric formulation as a standard Galerkin-based numerical approach \cite{Yiming:11}.  This work is embedded in the framework of the recently presented Global Cracking Elements Method \cite{Yiming:20}.  The implementation of the arc-length method is simple.  In this Section, only a brief introduction to this topic is provided.  Most of the symbols are the same as those used in \cite{Yiming:20}.
\subsection{Traction-separation law}
\begin{figure}[htbp]
	\centering
	\includegraphics[width=0.9\textwidth]{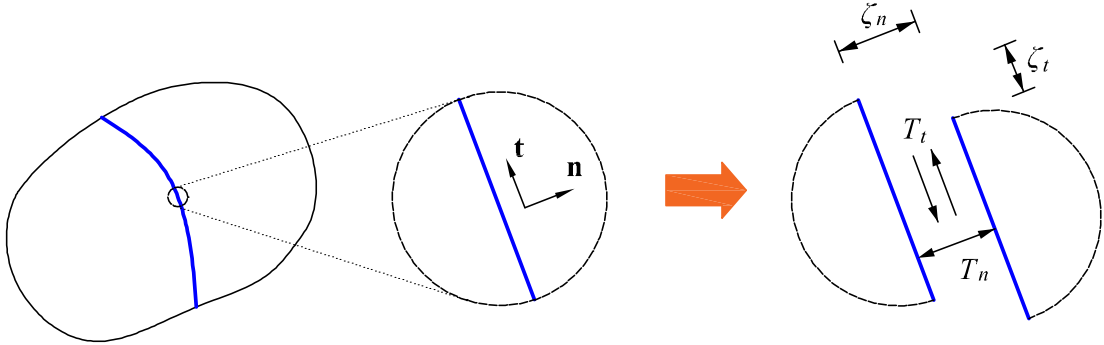}
	\caption{Definitions of $\mathbf{n}$, $\mathbf{t}$, $\left[\zeta_n,\ \zeta_t\right]$ and $\left[T_n,\ T_t\right]$}
	\label{fig:CrackDef}
\end{figure}

The mixed-mode traction-separation law \cite{Meschke:01,Yiming:15}, where the normal and the shear directions of the crack surface are defined by the unit vectors $\mathbf{n}=\left[n_x,\ n_y\right]^T$ and $\mathbf{t}=\left[t_x,\ t_y\right]^T$, is used in this work.  The crack openings along these two directions are denoted as $\boldsymbol{\zeta}=\left[\zeta_n,\ \zeta_t\right]^T$, and the traction across the two crack surfaces is denoted as $\mathbf{T}=\left[T_n,\ T_t\right]^T$; see Figure~\ref{fig:CrackDef}.

Based on $\boldsymbol{\zeta}$, $\mathbf{T}$ is obtained as
\begin{equation}
\begin{aligned}
&\left[\begin{array}{c}
T_n\\
T_t
\end{array}\right]=\frac{T_{eq}}{\zeta_{eq}}
\left[\begin{array}{c}
\zeta_n\\
\zeta_t
\end{array}\right],\\
&\mbox{with}\\
&\zeta_{eq}=\sqrt{\zeta_n^2+\zeta_t^2}\\
&\mbox{and }\\
&T_{eq}\left(\zeta_{eq} \right)=\left\{\begin{array}{ll}
L_1\left(\zeta_{eq} \right)=\cfrac{f_t}{\zeta_{0}}\ \zeta_{eq}& \mbox{for loading } \zeta_{eq}\leq \zeta_0,\\
L_2\left(\zeta_{eq} \right)=f_t \ \mbox{exp}\left[-\cfrac{f_t\left(\zeta_{eq}-\zeta_0\right)}{G_f-G_{f,0}}\right]& \mbox{for loading } \zeta_{eq}>\zeta_0,\\
U\left(\zeta_{eq} \right)=\cfrac{T_{mx}}{\zeta_{mx}}\ \zeta_{eq}& \mbox{for unloading/reloading}.
\end{array}\right.
\end{aligned}
\label{eq:Traction}
\end{equation}

In Eq.~\ref{eq:Traction}, $f_t$ denotes the uniaxial tensile strength and $G_f$ stands for the fracture energy.  $G_{f,0}$ denotes the threshold value of $G_f$, under the assumption of $G_{f,0}=0.01\ G_f$.  $\zeta_0$ is the corresponding threshold opening, with $\zeta_0=2\  G_{f,0}/f_t$.  $\zeta_{mx}$ stands for the maximum opening that the crack has ever experienced.  This value is updated at the end of each load step if $\zeta_{mx}>\zeta_0$.  $T_{mx}=L_2\left(\zeta_{mx}\right) $ is the corresponding traction.  Additional details can be found in \cite{Yiming:20}.  

Consequently, $\mathbf{D}=\partial \mathbf{T}\ / \ \partial \boldsymbol{\zeta}$ is obtained as
\begin{equation}
\begin{aligned}
&\mathbf{D}=\left[\begin{array}{cc}
{\partial T_n}/{\partial \zeta_n}&{\partial T_n}/{\partial \zeta_t}\\
{\partial T_t}/{\partial \zeta_n}&{\partial T_t}/{\partial \zeta_t}\\
\end{array}\right]=\\
&\left\{\begin{array}{ll}
\cfrac{f_t}{\zeta_0}\left[\begin{array}{cc}
1&0\\
0&1\\
\end{array}\right]& \mbox{for loading } \zeta_{eq}\leq \zeta_0,\\
\\
-\cfrac{T_{eq}}{\zeta_{eq}^2}\left[\begin{array}{cc}
\cfrac{\zeta_n^2}{\zeta_{eq}}+\cfrac{f_t\ \zeta_n^2}{{G_f-G_{f,0}}}-\zeta_{eq}&
\cfrac{\zeta_n\ \zeta_t}{\zeta_{eq}}+\cfrac{f_t\ \zeta_n\ \zeta_t}{{G_f-G_{f,0}}}\\
\cfrac{\zeta_n\ \zeta_t}{\zeta_{eq}}+\cfrac{f_t\ \zeta_n\ \zeta_t}{{G_f-G_{f,0}}}&
\cfrac{\zeta_t^2}{\zeta_{eq}}+\cfrac{f_t\ \zeta_t^2}{{G_f-G_{f,0}}}-\zeta_{eq}\\
\end{array}\right]& \mbox{for loading } \zeta_{eq}>\zeta_0,\\
\\
\cfrac{T_{mx}}{\zeta_{mx}}\left[\begin{array}{cc}
1&0\\
0&1\\
\end{array}\right]& \mbox{for unloading/reloading}.
\end{array}\right.
\label{eq:dTraction}
\end{aligned}
\end{equation}

This model is consistent with the conventional cohesive zone model \cite{Yiming:19}.  Herein, only an exponential-type law is used, but other types of traction-separation laws, such as linear, bilinear, and hyperbolic, can also be implemented.
\subsection{Kinematics and the global formulation}
\begin{figure}[htbp]
	\centering
	\includegraphics[width=0.9\textwidth]{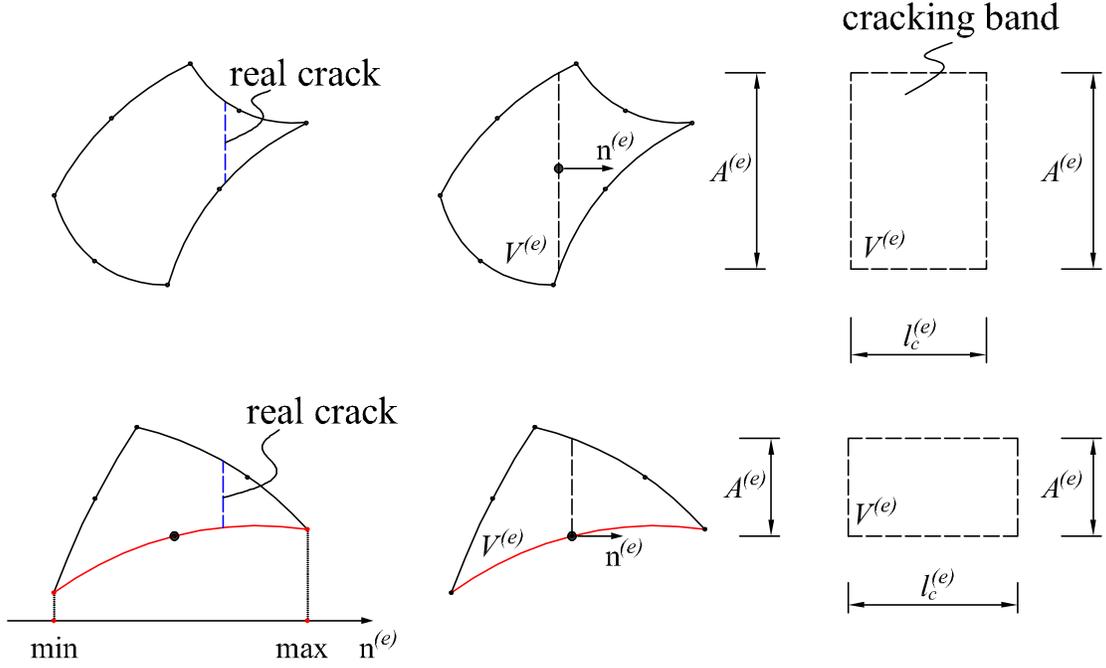}
	\caption{Relationships between $l_c$, $V^{(e)}$ and $A^{(e)}$ of Q8 and T6}
	\label{fig:lc}
\end{figure}

The notion of enhanced assumed strains (EAS) \cite{Simo:04} is used in the CEM.  Determination of $\mathbf{n}$ and $\mathbf{t}$ is not related to the formulation of the CEM, and these vectors are assumed to be known in this Section.  The total strain ${\boldsymbol{\varepsilon}}$ in the domain $\Omega\left( \mathbf{x} \right)$, experiencing cracking, consists of the elastic strain $\bar{\boldsymbol{\varepsilon}}$ and the enhanced strain $\tilde{\boldsymbol{\varepsilon}}$, i.e.
\begin{equation}
\begin{array}{ccccc}
\underbrace{\widehat{\boldsymbol{\varepsilon}}(\mathbf{x})=\nabla^S \bar{\mathbf{u}}(\mathbf{x})}&=  &\underbrace{\bar{\boldsymbol{\varepsilon}}(\mathbf{x})}  &+&\underbrace{\left[(\mathbf{n} \otimes \nabla \varphi)^S \zeta_n(\mathbf{x})+(\mathbf{t} \otimes \nabla \varphi)^S \zeta_t(\mathbf{x})\right]},\\
\mbox{total strain}&&\mbox{elastic strain}&&\mbox{enhanced strain } \widetilde{\boldsymbol{\varepsilon}}
\end{array}
\label{eq:EAS}
\end{equation}
where $\nabla \varphi$ is a vector of dimension ``length$^{-1}$" that links the crack openings to the enhanced strain.  $\nabla \varphi={\mathbf{n}}\ / \ {l_c}$ \cite{Yiming:11}, where $l_c$ corresponds to the classic characteristic length \cite{Oliver:02,Cervera:10}.  In the framework of the finite element method (FEM), Eq.~\ref{eq:EAS} yields
\begin{equation}
\begin{array}{cccc}
\bar{\boldsymbol{\varepsilon}}^{(e)}\approx&\underbrace{\sum^{n}_{i=1}\left(\nabla N^{(e)}_i \otimes \mathbf{u}_i\right)^S}&-&\underbrace{\frac{1}{\ l_c^{(e)} \ }\left[(\mathbf{n}^{(e)} \otimes  {\mathbf{n}}^{(e)}) \zeta_n^{(e)}+(\mathbf{n}^{(e)} \otimes  \mathbf{t}^{(e)})^S \zeta^{(e)}_t\right]},\\
&\widehat{\boldsymbol{\varepsilon}}^{(e)}&&\widetilde{\boldsymbol{\varepsilon}}^{(e)}
\end{array}
\label{eq:EAS_e}
\end{equation}
where $\left( \cdot \right)^S$ denotes the symmetric part of the tensor \cite{Mosler:01}, while $\left( \cdot \right)^{(e)}$ refers to the calculation of element $e$ to the respective quantity.  $n$ is the number of nodes  of a finite element.  The CEM has been proven to be reliable for implementations with 8-node quadrilateral elements (Q8, $n$=8) \cite{Yiming:14,Yiming:16} and 6-node triangular elements (T6, $n$=6) \cite{Yiming:21}.  Based on the conservation of energy, the element-dependent $l_c^{(e)}$ is obtained as $l_c^{(e)}=V^{(e)}\ / \ A^{(e)}$, where $V^{(e)}$ denotes the volume of element $e$ and $A^{(e)}$ stands for the surface area of an equivalent crack parallel to the real crack.  Here, the determination of $A^{(e)}$ for Q8 and T6 is slightly different insofar as the equivalent crack passes through the center point of Q8 but through the midpoint of one edge of T6; see Figure~\ref{fig:lc}.  More details can be found in \cite{Yiming:20,Yiming:21}.

One static and one kinematic balance equation hold for the cracking element.  For the quasi-static loading condition, their quantities are given as
\begin{equation}
\begin{aligned}
&\nabla \boldsymbol{\sigma}-\mathbf{F}=\mathbf{0},\\
&\mbox{and}\\
&\left[ {\begin{array}{c}
	\mathbf{n}^{(e)}\otimes\mathbf{n}^{(e)}\\
	\mathbf{n}^{(e)}\otimes\mathbf{t}^{(e)}
	\end{array} } \right]:\boldsymbol{\sigma}^{(e)}-\mathbf{T}^{(e)}=\mathbf{0}, \forall \ e,
\label{eq:equcenter}
\end{aligned}
\end{equation}
where $\mathbf{F}$ denotes the loading force and  $\boldsymbol{\sigma}^{(e)}=\mathbb{C}^{(e)}:\bar{\boldsymbol{\varepsilon}}^{(e)}$, with $\mathbb{C}^{(e)}$ as the elasticity tensor.  

The authors of \cite{Yiming:20} reformulated the CEM in matrix form, which is very simple and convenient.  The symmetric second- and fourth-order tensors are represented by vector and matrix forms, respectively, as Voigt's notation \cite{Heinwein:01}.  The displacement vector is given as $\mathbf{U}^{(e)}=\left[\mathbf{u}^{(e)}_1 \cdots \mathbf{u}^{(e)}_n\right]^T$.  The total strain is obtained as $\widehat{\boldsymbol{\varepsilon}}^{(e)}=\mathbf{B}^{(e)}\mathbf{U}^{(e)}$, where the $\mathbf{B}^{(e)}$ matrix is defined as
\begin{equation}
\begin{aligned}
&\mathbf{B}^{(e)}=
\left[ {\begin{array}{ccc}
	\mathbf{B}^{(e)}_1&\cdots&\mathbf{B}^{(e)}_n\\
	\end{array} } \right],\\
&\mbox{with}\\
&\mathbf{B}^{(e)}_i=\left[\begin{array}{ccc}
\cfrac{\partial N_i^{(e)}}{\partial x} &0\\
0&\cfrac{\partial N_i^{(e)}}{\partial y} \\
\cfrac{\partial N_i^{(e)}}{\partial y}&\cfrac{\partial N_i^{(e)}}{\partial x}
\end{array}
\right],\ i=1 \cdots n.\\
\label{eq:B}
\end{aligned}
\end{equation} 
Then, based on Eq.~\ref{eq:EAS_e}, the following matrix is introduced:
\begin{equation}
\mathbf{B}^{(e)}_\zeta=\frac{-1}{\ l_c^{(e)} \ }
\left[ {\begin{array}{c}
	\mathbf{n}^{(e)}\otimes\mathbf{n}^{(e)}\\
	\left(\mathbf{n}^{(e)}\otimes\mathbf{t}^{(e)}\right)^S
	\end{array} } \right]^T
=\frac{-1}{\ l_c^{(e)} \ }
\left[ {\begin{array}{cc}
	n^{(e)}_x\cdot n^{(e)}_x&n^{(e)}_x\cdot t^{(e)}_x\\
	n^{(e)}_y\cdot n^{(e)}_y&n^{(e)}_y\cdot t^{(e)}_y\\
	2\ n^{(e)}_x\cdot n^{(e)}_y&n_x\cdot t^{(e)}_y+n^{(e)}_y\cdot t^{(e)}_x
	\end{array} } \right].
\label{eq:Bz}
\end{equation} 
Considering $\left(\mathbf{n}^{(e)}\otimes\mathbf{t}^{(e)}\right)^S:\boldsymbol{\sigma}=\left(\mathbf{n}^{(e)}\otimes\mathbf{t}^{(e)}\right):\boldsymbol{\sigma}$.  By using the matrices $\mathbf{B}^{(e)}$ and $\mathbf{B}^{(e)}_\zeta$, after considering the weak form, Eq.~\ref{eq:equcenter} is rewritten as 
\begin{equation}
\begin{aligned}
&\int \left[ \left( \mathbf{B}^{(e)}\right)^T \ \mathbf{C}^e \ \bar{\boldsymbol{\varepsilon}}^{(e)} - \mathbf{F}^{(e)}\right] d(e)=\mathbf{0},\\
&\mbox{and}\\
&-l_c\left(\mathbf{B}^{(e)}_\zeta\right)^T\boldsymbol{\sigma}^{(e)}-\left[ {\begin{array}{c}
	T_n^{(e)}\\
	T_t^{(e)}
	\end{array} } \right]=\mathbf{0}, \forall \ e,
\label{eq:equcentermatrix}
\end{aligned}
\end{equation}
where $\mathbf{C}^e$ denotes the matrix form of $\mathbb{C}^{(e)}$.

Meanwhile, Eq.~\ref{eq:EAS_e} gives
\begin{equation}
\bar{\boldsymbol{\varepsilon}}^{(e)}=\bar{\boldsymbol{\varepsilon}}^{(e),1}=\left[ {\begin{array}{cc}
	\mathbf{B}^{(e),1}&	\mathbf{B}^{(e)}_\zeta
	\end{array} } \right] \left[ {\begin{array}{c}
	\mathbf{U}^{(e)}\\
	\boldsymbol{\zeta}^{(e)}
	\end{array} } \right].
\label{eq:EASQ8}
\end{equation}
where $\mathbf{B}^{(e),1}$ denotes $\mathbf{B}^{(e)}$ evaluated at the center Gauss point (center representation) \cite{Yiming:20}.

According to the Newton-Raphson (N-R) method, for the iteration step $l$ at load step $i$ the element-related incremental relation is obtained as follows:
\begin{equation}
\begin{array}{cccc}
\left[ {\begin{array}{c}
	\mathbf{U}_{i,l}^{(e)}\\
	\boldsymbol{\zeta}_{i,l}^{(e)}
	\end{array} } \right]=
&\underbrace{
	\left[ {\begin{array}{c}
		\mathbf{U}^{(e)}_{i-1}\\\boldsymbol{\zeta}^{(e)}_{i-1}
		\end{array} } \right]+
	\left[ {\begin{array}{c}
		\Delta\mathbf{U}^{(e)}_{l-1}\\\Delta\boldsymbol{\zeta}^{(e)}_{l-1}
		\end{array} } \right]}&+&\underbrace{
	\left[ {\begin{array}{c}
		\Delta\Delta\mathbf{U}^{(e)}\\\Delta\Delta\boldsymbol{\zeta}^{(e)}
		\end{array} } \right]}.\\
&\mbox{known}&&\mbox{unknown}\\
\end{array}
\label{eq:UdU2}
\end{equation}

In Eq.~\ref{eq:UdU2}, $\Delta\left(\cdot\right)$ denotes an increment of the corresponding value at the preceding load step, $i-1$, while $\Delta\Delta\left(\cdot\right)$ stands for an increment of the value at the last N-R iteration step, $l-1$.

After linearization, the element-related balance equation, deduced in \cite{Yiming:20}, is obtained as follows:
\begin{equation}
\begin{aligned}
&\mathbf{K}_{sym}^{(e)}
\left[ {\begin{array}{c}
	\Delta\Delta\mathbf{U}^{(e)}\\\Delta\Delta\boldsymbol{\zeta}^{(e)}
	\end{array} } \right]=
\left[ {\begin{array}{c}
	\mathbf{F}\\-\cfrac{V^{(e)}}{l_c^{(e)}}\ 
 \mathbf{T}^{(e)}
	\end{array} } \right]-\mathbf{K}^{(e)}
\left[ {\begin{array}{c}
	\mathbf{U}^{(e)}_{i-1}+\Delta\mathbf{U}^{(e)}_{l-1}\\\boldsymbol{\zeta}^{(e)}_{i-1}+\Delta\boldsymbol{\zeta}^{(e)}_{l-1}
	\end{array} } \right],\\
&\mbox{with}\\
&\mathbf{K}^{(e)}=
\left[ {\begin{array}{cc}
	\int \left(\mathbf{B}^{(e)}\right)^T \mathbf{C}^{(e)} \ \left(\mathbf{B}^{(e),1}\right) d(e) &\int \left(\mathbf{B}^{(e)}\right)^T \mathbf{C}^{(e)} \ \mathbf{B}_\zeta^{(e)} d(e) \\
	V^{(e)} \left(\mathbf{B}^{(e)}_\zeta\right)^T\mathbf{C}^{(e)} \ \left(\mathbf{B}^{(e),1}\right) &V^{(e)} \left(\mathbf{B}^{(e)}_\zeta\right)^T\mathbf{C}^{(e)} \ \mathbf{B}^{(e)}_\zeta\\
	\end{array} } \right]\\
&\mbox{and}\\
&\mathbf{K}_{sym}^{(e)}=\int \left[ {\begin{array}{cc}
	\mathbf{B}^{(e)}&	\mathbf{B}^{(e)}_\zeta
	\end{array} } \right]^T \mathbf{C}^{(e)} \left[ {\begin{array}{cc}
	\mathbf{B}^{(e)}&	\mathbf{B}^{(e)}_\zeta
	\end{array} } \right] d(e) +
\left[ {\begin{array}{cc}
	\mathbf{0}&\mathbf{0}\\
	\mathbf{0}&\cfrac{V^{(e)}}{l_c^{(e)}} \ \mathbf{D}^{(e)}\\
	\end{array} } \right].
\label{eq:Ksub2}
\end{aligned}
\end{equation}
While $\mathbf{K}^{(e)}$ is an unsymmetric matrix with many zero elements, $\mathbf{K}_{sym}^{(e)}$ is a symmetric and positive definite matrix.  In this work a dynamic element-related enrichment version of the CEM is used \cite{WU2015346}.  Once element $e$ experiences cracking, the additional degree of freedom, $\boldsymbol{\zeta}^{(e)}$, is introduced as a new global unknown.  
\subsection{Crack propagation}
A distinguishing feature of the CEM from its standard Galerkin form is its self-propagating crack.  As indicated in the previous Section, the CEM does not need continuous crack paths.  Instead, only the local orientation $\mathbf{n}$ of the crack needs to be known, making a local criterion simple, efficient, and robust.  

As a local criterion, $\mathbf{n}^{(e)}$ depends only on $\widehat{\boldsymbol{\varepsilon}}^{(e)}$ and is not related to any other element.  It is assumed to be the first unit eigenvector of the total strain $\widehat{\boldsymbol{\varepsilon}}$ at the center point, i.e.,
\begin{equation}
\begin{aligned}
&\widehat{\boldsymbol{\varepsilon}}^{(e)}\cdot\mathbf{n}^{(e)}-\widehat{{\varepsilon}}_1^{(e)}\cdot\mathbf{n}^{(e)}=\mathbf{0},\\
&\mbox{where}\\
&\widehat{\boldsymbol{\varepsilon}}^{(e)}=\sum^{n}\limits_{i=1}\left(\nabla N^{(e)}_i \otimes \mathbf{u}_i\right)^S=\left[\begin{array}{cc}
\widehat{\varepsilon}_x^{(e)}&\widehat{\gamma}_{xy}^{(e)}\ / \ 2\\
\widehat{\gamma}_{xy}^{(e)}\ / \ 2&\widehat{\varepsilon}_y^{(e)} \end{array}\right],\\
&\mbox{and}\\
&\widehat{{\varepsilon}}_1^{(e)}=\frac{\widehat{{\varepsilon}}_x^{(e)}+\widehat{{\varepsilon}}_y^{(e)}+\sqrt{\left(\widehat{{\varepsilon}}_x^{(e)}-\widehat{{\varepsilon}}_y^{(e)} \right)^2+\left(\widehat{\gamma}_{xy}^{(e)}\right)^2}}{2}.
\label{eq:local}
\end{aligned}
\end{equation}

The elements experience cracking, one after another: crack propagation is always checked first; then, crack initiation is considered. The following strategy is used to identify the next cracking element:
\begin{equation}
\begin{aligned}
&\mbox{find }\ \mbox{max}\left\{\phi_{RK}^{(e)}\right\} \mbox{ with} \\
&\phi_{RK}^{(e)}=\left(\mathbf{n}^{(e)}\otimes\mathbf{n}^{(e)}\right):\mathbb{C}^{(e)}:\widehat{\boldsymbol{\varepsilon}}^{(e)}-f_t^{(e)}\\
&\mbox{and }\phi_{RK}^{(e)}>0, \nonumber
\label{eq:phiRK}
\end{aligned}
\end{equation}
where $\phi_{RK}^{(e)}>0$ can be considered as a Rankine-like criterion.  If $\phi_{RK}^{(e)}<0$ for all non-cracked elements, the iteration stops.  This strategy was first presented in \cite{Yiming:14}, and a detailed flowchart was provided in \cite{Yiming:20}.

\section{Direct dissipation-based arc-length approach}
\label{sec:arc}
\subsection{Basics}
For a deformed body with cohesive discontinuities, the mechanical energy \cite{xie:01,Meschke:01} $\Psi$ is given as
\begin{equation}
\Psi=I+E-W,
\label{eq:psi1}
\end{equation}
where $I$ is the elastic strain energy (internal energy), $E$ is the dissipated energy, and $W$ is the work done by the applied forces.  

$\Psi=0$ is obtained for a conservative system; hence, most energy-based arc-length methods use $W-I$ and not $E$ as the state variable \cite{Verhoosel2008}.  This is reasonable since $W-I$ is related to the displacements and $E$ is related to crack openings.  However, since cracks are propagating, $\Psi$ is a function of the crack direction and surface area.  Both are changing continuously during the loading process \cite{xie:01,Dumstorff:01}.  For $\Psi>0$, $W-I<0$ may occur.  In this case, $W-I$ cannot be used as the state variable.  Moreover, because of the numerical error in the iteration step, the difference between $W-I$ and $E$ is further increased.  This difference is more obvious for crack-opening based models than damage-degree based models, because crack openings are reversible and without upper bounds, whereas the damage degree is irreversible and bounded. 

The CEM is a crack-opening based model \cite{Yiming:19}.  For better numerical stability, $E$ is used directly as the state variable of the arc-length method.  For convenience, the following global vectors are defined:

\begin{equation}\boldsymbol{\mathsf{Z}}=\bigcup \boldsymbol{\zeta}^{(e)}\ \mbox{ and }\ \frac{\ V\ }{\ l_c\ } \boldsymbol{\mathsf{T}}=\bigcup\left(\frac{V^{(e)}}{l_c^{(e)}}\ \mathbf{T}^{(e)}\right),
\label{eq:vectorg}
\end{equation}
where $\bigcup\left(\cdot\right)$ denotes the assemblage of the element matrix or vector to the global form.  Then, considering Eqs.~\ref{eq:UdU2} and~\ref{eq:Ksub2}, by means of a forward Euler discretization, the incremental dissipated energy at load step $i$ is obtained as 
\begin{equation}
\begin{aligned}
&d E_i=\frac{\ 1 \ }{\ 2 \ }\left[\Delta \boldsymbol{\mathsf{Z}}_i\ \left(\frac{\ V\ }{\ l_c\ } \mathsf{T}^T\right)_{i-1}+\boldsymbol{\mathsf{Z}}_{i-1}\ \Delta \left(\frac{\ V\ }{\ l_c\ } \mathsf{T}^T\right)_i\right]\overset{!}{=}a,\\
&\mbox{where}\\
&\Delta \left(\frac{\ V\ }{\ l_c\ } \mathsf{T}^T\right)_i=\left(\frac{\ V\ }{\ l_c\ } \mathsf{T}^T\right)_{i}-\left(\frac{\ V\ }{\ l_c\ } \mathsf{T}^T\right)_{i-1},
\end{aligned}
\label{eq:deltaE}
\end{equation}
which is used as the arc-length restraint, where $a$ is the prescribed arc-length.  $a$ is determined at the beginning of every load step, based on the total residual dissipated energy of the system.  In general, $a$ is taken as 1$\%$ of the total residual dissipated energy.
\subsection{Formulation}
Similar to Eq.~\ref{eq:vectorg}, the following global matrices and vectors are defined:
\begin{equation}
\begin{aligned}
&\boldsymbol{\mathsf{K}}_{sym}=\bigcup\mathbf{K}_{sym}^{(e)}\ \mbox{ and }\ \boldsymbol{\mathsf{K}}=\bigcup\mathbf{K}^{(e)}\\
&\boldsymbol{\mathsf{U}}=\bigcup\mathbf{U}^{(e)}\ \mbox{ and }\ \boldsymbol{\mathsf{F}}=\bigcup\mathbf{F}^{(e)}=\lambda \ \mathsf{f},
\label{eq:vectorgm}
\end{aligned}
\end{equation}
where $\lambda$ is the unknown arc-length for controlling the load and $\mathsf{f}$ is a prescribed reference load.  Based on Eq.~\ref{eq:UdU2}, for iteration step $l$ at load step $i$, the global arc-length based incremental relation is obtained as 
\begin{equation}
\begin{array}{cccc}
\left[ \begin{array}{c}
	\boldsymbol{\mathsf{U}}_{i,l}\\ \boldsymbol{\mathsf{Z}}_{i,l}\\ \lambda_{i,l}
	\end{array}  \right]=
&\underbrace{
	\left[ {\begin{array}{c}
		\boldsymbol{\mathsf{U}}_{i-1}\\\boldsymbol{\mathsf{Z}}_{i-1}\\ \lambda_{i-1}
		\end{array} } \right]+
	\left[ {\begin{array}{c}
		\Delta\boldsymbol{\mathsf{U}}_{l-1}\\\Delta\boldsymbol{\mathsf{Z}}_{l-1}\\\Delta \lambda_{l-1}
		\end{array} } \right]}&+&\underbrace{
	\left[ {\begin{array}{c}
		\Delta\Delta\boldsymbol{\mathsf{U}}\\\Delta\Delta\boldsymbol{\mathsf{Z}}\\\Delta\Delta\lambda
		\end{array} } \right]}.\\
&\mbox{known}&&\mbox{unknown}\\
\end{array}
\label{eq:UdUg}
\end{equation}
Correspondingly, for iteration step $l$ at load step $i$ if $\Delta\Delta\lambda\rightarrow0$, after assembling the element balance Eq.~\ref{eq:Ksub2}, the following global balance equation is obtained:
\begin{equation}
\boldsymbol{\mathsf{K}}_{sym}
	\left[ {\begin{array}{c}
		\Delta\Delta\boldsymbol{\mathsf{U}}\\\Delta\Delta\boldsymbol{\mathsf{Z}}
		\end{array} } \right]=\left[ {\begin{array}{c}
		\left(\lambda_{i-1}+\Delta \lambda_{l-1}\right)\mathsf{f}\\
		-\left(\cfrac{\ V\ }{\ l_c\ }\  \mathsf{T}\right)_{l-1}
		\end{array} } \right]-\boldsymbol{\mathsf{K}}
	\left[ {\begin{array}{c}
		\boldsymbol{\mathsf{U}}_{i-1}+\Delta \boldsymbol{\mathsf{U}}_{l-1}\\
		\boldsymbol{\mathsf{Z}}_{i-1}+\Delta \boldsymbol{\mathsf{Z}}_{l-1}
		\end{array} } \right].
\label{eq:Ksub2new}
\end{equation}

When considering $\Delta\Delta\lambda$ and the restraint Eq.\ref{eq:deltaE}, the global balance equation is obtained as
\begin{equation}
\begin{aligned}
&\left[
	\begin{array}{cc:c}
	\multicolumn{2}{c:}{\multirow{2}*{$\boldsymbol{\mathsf{K}}_{sym}$}}&-\mathsf{f}\\
	& &\mathbf{0}\\ 
	\hdashline
	\mathbf{0}&\left(\cfrac{\ V\ }{\ l_c\ }\  \mathsf{T}^T\right)_{i-1}&k_\lambda
	\end{array}
	\right]
	\left[ {\begin{array}{c}
	 \Delta\Delta\boldsymbol{\mathsf{U}}\\\Delta\Delta\boldsymbol{\mathsf{Z}}\\
	 \hdashline\Delta\Delta\lambda
	\end{array} } \right]=	\left[ {\begin{array}{c}
	\mathsf{R_U}\\
	\mathsf{R_Z}\\
	\hdashline
	R_\lambda
	\end{array} } \right],\\
&\mbox{where}\\
&\left[ {\begin{array}{c}
	\mathsf{R_U}\\
	\mathsf{R_Z}\\
	\hdashline
	R_\lambda
	\end{array} } \right]=\left[ {\begin{array}{c}
	\left(\lambda_{i-1}+\Delta \lambda_{l-1}\right)\mathsf{f}\\
	-\left(\cfrac{\ V\ }{\ l_c\ }\  \mathsf{T}\right)_{l-1}\\
	\hdashline
	2\ a
	\end{array} } \right]-
\left[ {\begin{array}{c:c}
	\boldsymbol{\mathsf{K}}&1
	\end{array} } \right]
\left[ {\begin{array}{c}
	\boldsymbol{\mathsf{U}}_{i-1}+\Delta \boldsymbol{\mathsf{U}}_{l-1}\\
	\boldsymbol{\mathsf{Z}}_{i-1}+\Delta \boldsymbol{\mathsf{Z}}_{l-1}\\
	\hdashline
	2\ d E_{l-1}
	\end{array} } \right],\\
&\mbox{with}\\
& d E_{l-1}=\frac{\ 1 \ }{\ 2 \ }\left[\Delta \boldsymbol{\mathsf{Z}}_{l-1} \left(\frac{\ V\ }{\ l_c\ } \mathsf{T}^T\right)_{i-1}+\boldsymbol{\mathsf{Z}}_{i-1} \Delta \left(\frac{\ V\ }{\ l_c\ } \mathsf{T}^T\right)_{l-1}\right]\\
&\mbox{and}\\
&\Delta \left(\frac{\ V\ }{\ l_c\ } \mathsf{T}^T\right)_{l-1}=\left(\frac{\ V\ }{\ l_c\ } \mathsf{T}^T\right)_{l-1}-\left(\frac{\ V\ }{\ l_c\ } \mathsf{T}^T\right)_{i-1}.
\label{eq:globalK}
\end{aligned}
\end{equation}
This equation raises two concerns: 
\begin{itemize}
	\item 
	the coefficient matrix is unsymmetric;
	\item
	$k_\lambda$ is unknown.
\end{itemize}
In the next Section, it will be shown that both concerns can be refuted by the Sherman-Morrison formula.

\subsection{Sherman-Morrison formula}
The Sherman-Morrison formula, which is widely used in arc-length methods \cite{Verhoosel2008}, separates the arc-length equation from the global equation to enable a more efficient solution.  Based on this formula, two vectors are obtained by solving
\begin{equation}
\boldsymbol{\mathsf{K}}_{sym}
\left[ {\begin{array}{cc}
	\Delta\Delta\boldsymbol{\mathsf{U}}_I&\Delta\Delta\boldsymbol{\mathsf{U}}_{II}\\
	\Delta\Delta\boldsymbol{\mathsf{Z}}_I&\Delta\Delta\boldsymbol{\mathsf{Z}}_{II}
	\end{array} } \right]=\left[ {\begin{array}{cc}
	\mathsf{R_U}&-\mathsf{f}\\
	\mathsf{R_Z}&\mathbf{0}\\
	\end{array} } \right].
\label{eq:sm1}
\end{equation}
This can be done by powerful linear solvers such as PARDISO \cite{pardiso,Kourounis2018} and MUMPS \cite{mumps,AMESTOY2000501,AMESTOY2003833}.  MUMPS 5.1.2 is used in this work.  The symmetry of the original system is maintained.  Thus, the first concern is refuted.

As regards the second concern, by definition, $k_\lambda$ is ``the change in $d E_i$ resulting from the unit increment $\mathsf{f}$ ".  Since $\left[\Delta\Delta\boldsymbol{\mathsf{U}}_{II}, \Delta\Delta\boldsymbol{\mathsf{Z}}_{II}\right]^T$ can be considered as the additional system response caused by $-\mathsf{f}$, $k_\lambda$ can be obtained by means of Eq.~\ref{eq:deltaE} as the difference between the trial values of $d E_i$.  This involves the following steps:
\begin{equation}
\begin{aligned}
&\mbox{obtain } dE_I\mbox{ from } \boldsymbol{\mathsf{Z}}_{i-1} \mbox{ and } \Delta\boldsymbol{\mathsf{Z}}_{l-1},\\
&\mbox{obtain } dE_{II}\mbox{ from } \boldsymbol{\mathsf{Z}}_{i-1} \mbox{ and } \Delta\boldsymbol{\mathsf{Z}}_{l-1}-\Delta\Delta\boldsymbol{\mathsf{Z}}_{II} ,\\
&\mbox{then compute }k_\lambda=dE_{II}-dE_I. \nonumber
\end{aligned}
\label{eq:klambda}
\end{equation}
Notably, the procedure attained above can be considered as a general strategy for obtaining $k_\lambda$, as used in most arc-length methods.  This procedure is especially suitable for cases where the derivative of the state variable with respect to $\lambda$ cannot be computed analytically.

Then, the final solution is obtained as
\begin{equation}
\begin{aligned}
&\left[ {\begin{array}{c}
	\Delta\Delta\boldsymbol{\mathsf{U}}\\
	\Delta\Delta\boldsymbol{\mathsf{Z}}\\
	\Delta\Delta\lambda
	\end{array} } \right]=
\left[ {\begin{array}{c}
	\Delta\Delta\boldsymbol{\mathsf{U}}_I\\
	\Delta\Delta\boldsymbol{\mathsf{Z}}_I\\
	R_\lambda
	\end{array} } \right]-\frac{1}{S_{II}-k_\lambda}
\left[ {\begin{array}{c}
	\left( S_I- R_\lambda \right)\Delta\Delta\boldsymbol{\mathsf{U}}_{II}\\
	\left( S_I- R_\lambda \right)\Delta\Delta\boldsymbol{\mathsf{Z}}_{II}\\
	-S_I+R_\lambda\left(1+S_{II}-k_\lambda\right)	
	\end{array} } \right],\\
&\mbox{where}\\
&S_I=\left[ {\begin{array}{cc}
	\mathbf{0}&\left(\cfrac{\ V\ }{\ l_c\ }\  \mathsf{T}^T\right)_{i-1}
	\end{array} } \right]\left[ {\begin{array}{c}
	\Delta\Delta\boldsymbol{\mathsf{U}}_I\\
	\Delta\Delta\boldsymbol{\mathsf{Z}}_I
	\end{array} } \right]\\
&\mbox{and}\\
&S_{II}=\left[ {\begin{array}{cc}
	\mathbf{0}&\left(\cfrac{\ V\ }{\ l_c\ }\  \mathsf{T}^T\right)_{i-1}
	\end{array} } \right]\left[ {\begin{array}{c}
	\Delta\Delta\boldsymbol{\mathsf{U}}_{II}\\
	\Delta\Delta\boldsymbol{\mathsf{Z}}_{II}
	\end{array} } \right],
\label{eq:sm3}
\end{aligned}
\end{equation}

\section{Numerical examples}
\label{sec:ne}
The plane stress condition is assumed for all numerical examples.
\subsection{Double-notched four-point bending test}
\begin{figure}[htbp]
	\centering
	\includegraphics[width=0.9\textwidth]{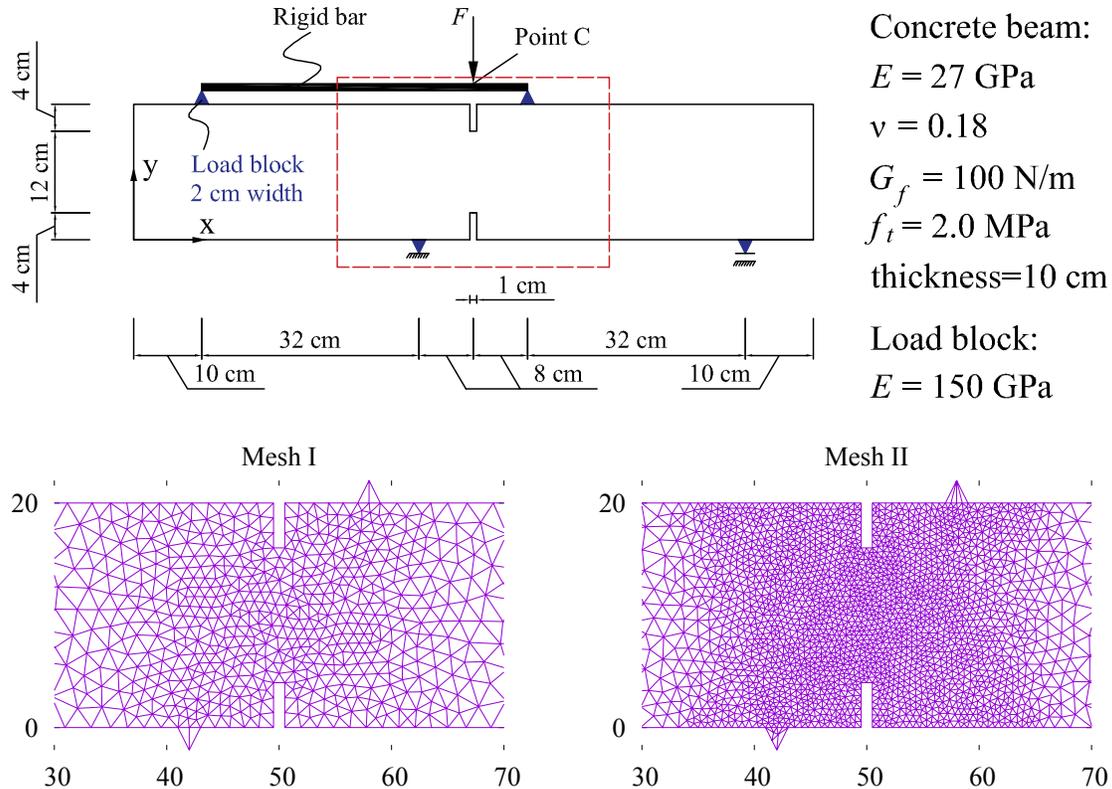}
	\caption{Double-notched four-point bending test: model, material, and meshes}
	\label{fig:AdenModel}
\end{figure}

The double-notched four-point bending test of a concrete beam \cite{Bocca:01}, illustrated in Figure~\ref{fig:AdenModel}, is a benchmark test investigated, e.g., in \cite{Oliver:09,Geers:01,Holzapfel:01}.  The load blocks in Figure~\ref{fig:AdenModel} are assumed to be unbreakable.  Because of symmetry, two axisymmetric cracks begin to propagate in the early stage of loading, but only one crack continuously grows until the beam fails.  From this perspective, the path is difficult to trace via crack mouth opening, since both crack mouths open in the early stage of loading but only one opens continuously while the other one closes.

\begin{figure}[htbp]
	\centering
	\includegraphics[width=0.95\textwidth]{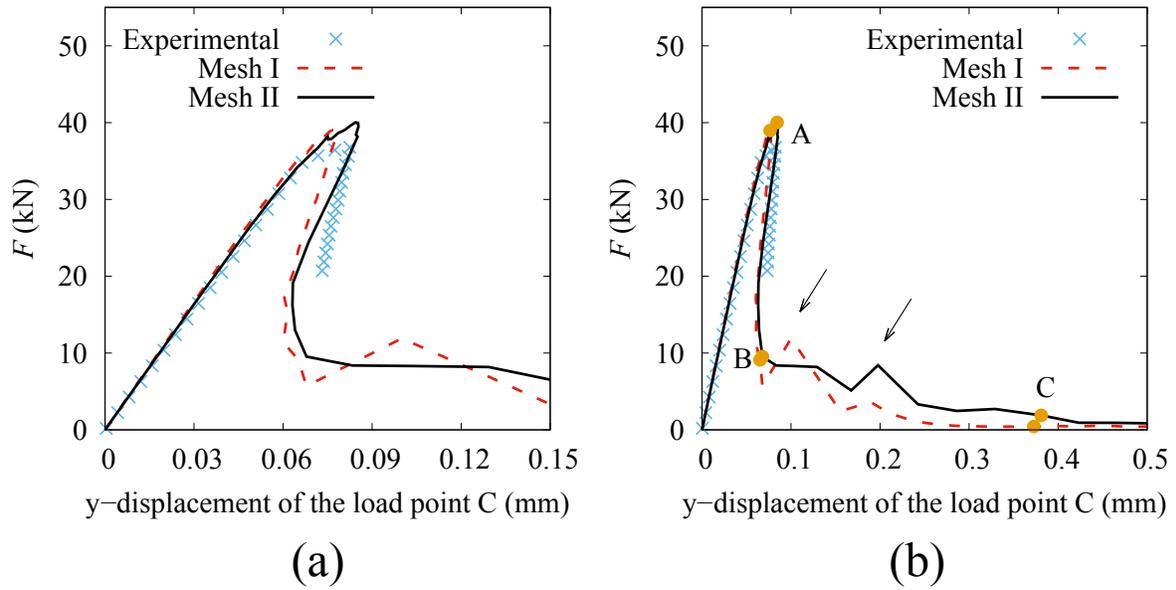}
	\caption{Double-notched four-point bending test: force-displacement curves (a) early stage, (b) numerical results compared to the experimental results given in \cite{Bocca:01}}
	\label{fig:AdenReF}
\end{figure}

\begin{figure}[htbp]
	\centering
	\includegraphics[width=0.95\textwidth]{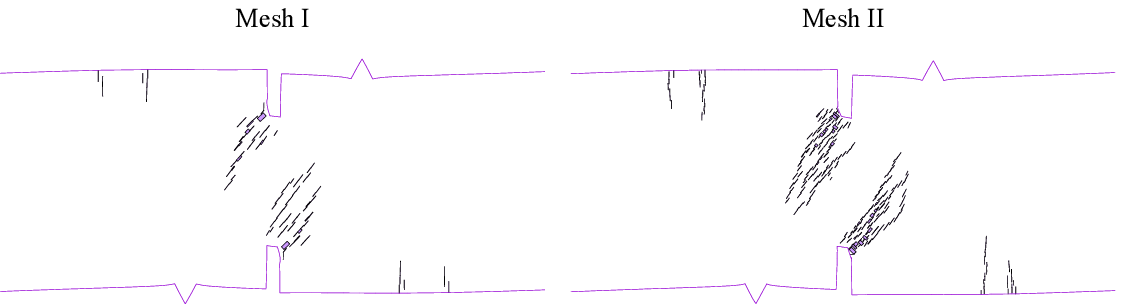}
	\caption{Double-notched four-point bending test: crack opening plots with deformed boundaries at stage A (scale = 1:200)}
	\label{fig:AdenReCW1}
\end{figure}

\begin{figure}[htbp]
	\centering
	\includegraphics[width=0.95\textwidth]{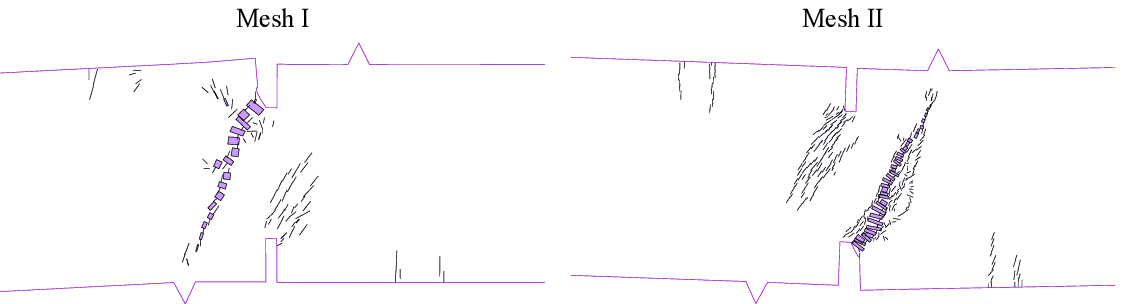}
	\caption{Double-notched four-point bending test: crack opening plots with deformed boundaries at stage B (scale = 1:50)}
	\label{fig:AdenReCW2}
\end{figure}

\begin{figure}[htbp]
	\centering
	\includegraphics[width=0.95\textwidth]{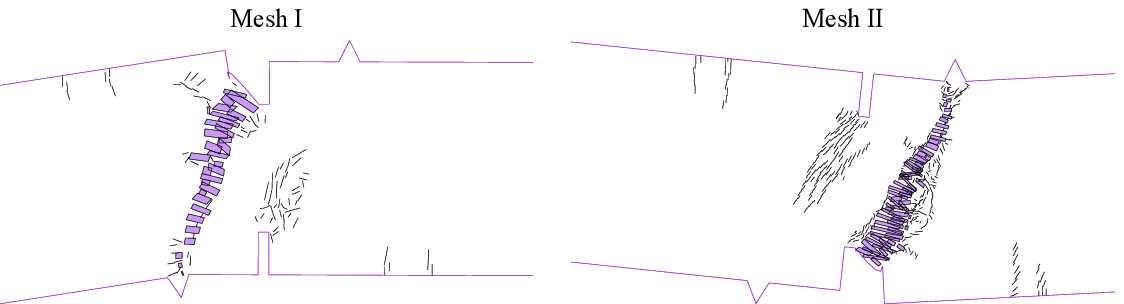}
	\caption{Double-notched four-point bending test: crack opening plots with deformed boundaries at stage C (scale = 1:20)}
	\label{fig:AdenReCW3}
\end{figure}

The force-displacement curves shown in Figure~\ref{fig:AdenReF} are compared to the experimental results given in \cite{Bocca:01}.  In Figure~\ref{fig:AdenReF}(b), there are two local peak loads marked by arrows.  They are caused by the approach of the cracks to the unbreakable load blocks.  The  crack opening plots with deformed boundaries at stages A, B, and C (see Figure~\ref{fig:AdenReF}(b)) are shown in Figures~\ref{fig:AdenReCW1} -~\ref{fig:AdenReCW3}.  As mentioned previously, two major cracks propagate, but only one grows continuously.  For both meshes, the dissipation-based arc-length approach captures the cracking process.

\subsection{Perforated plate with a hole}
\begin{figure}[htbp]
	\centering
	\includegraphics[width=0.7\textwidth]{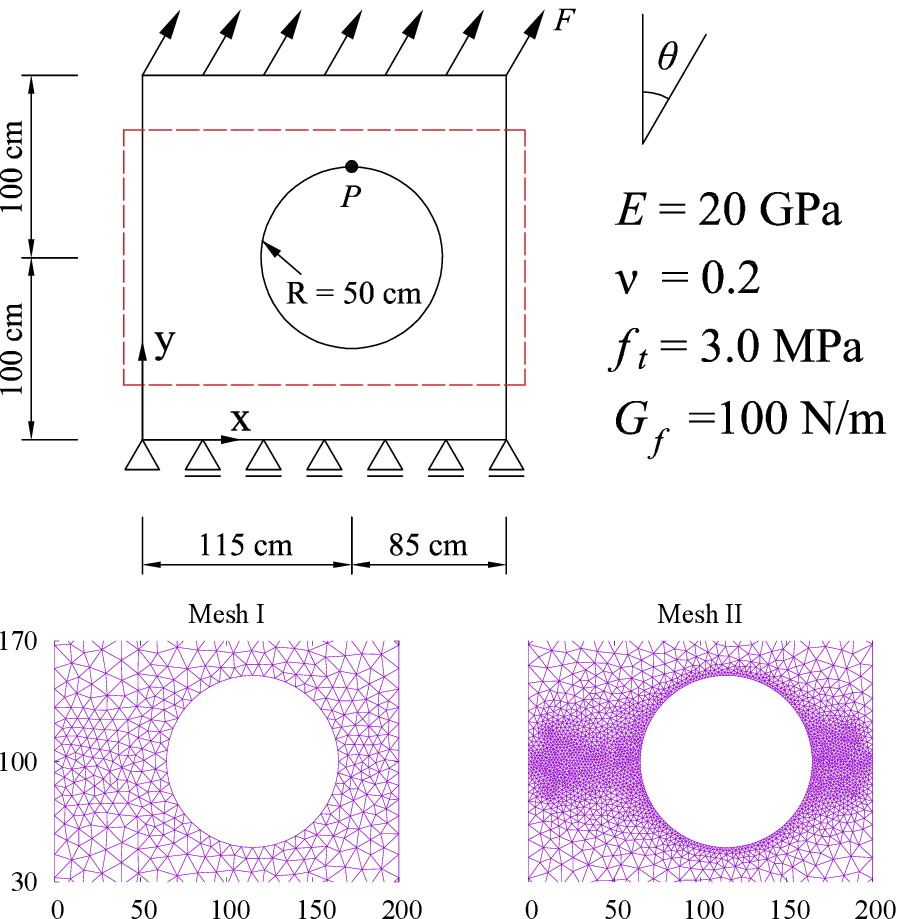}
	\caption{Perforated plate with a hole: model, materials, and meshes}
	\label{fig:RecHmodel}
\end{figure}

A perforated plate with a hole is a benchmark test for testing arc-length methods, such as in \cite{Lorentz2004,MEJIASANCHEZ2020405,LABANDA2018319,OZDEMIR2019208}.  This is an example for the absence of a crack mouth.  The model and the meshes are shown in Figure~\ref{fig:RecHmodel}.  The distributed force is applied at the top edge.  The bottom edge is fixed in the restricted direction.  In this example, neither an interface element nor a prescribed crack path is used: all cracks self-propagate during the arc-length loading process, which is a special advantage of the CEM.  

\begin{figure}[htbp]
	\centering
	\includegraphics[width=0.75\textwidth]{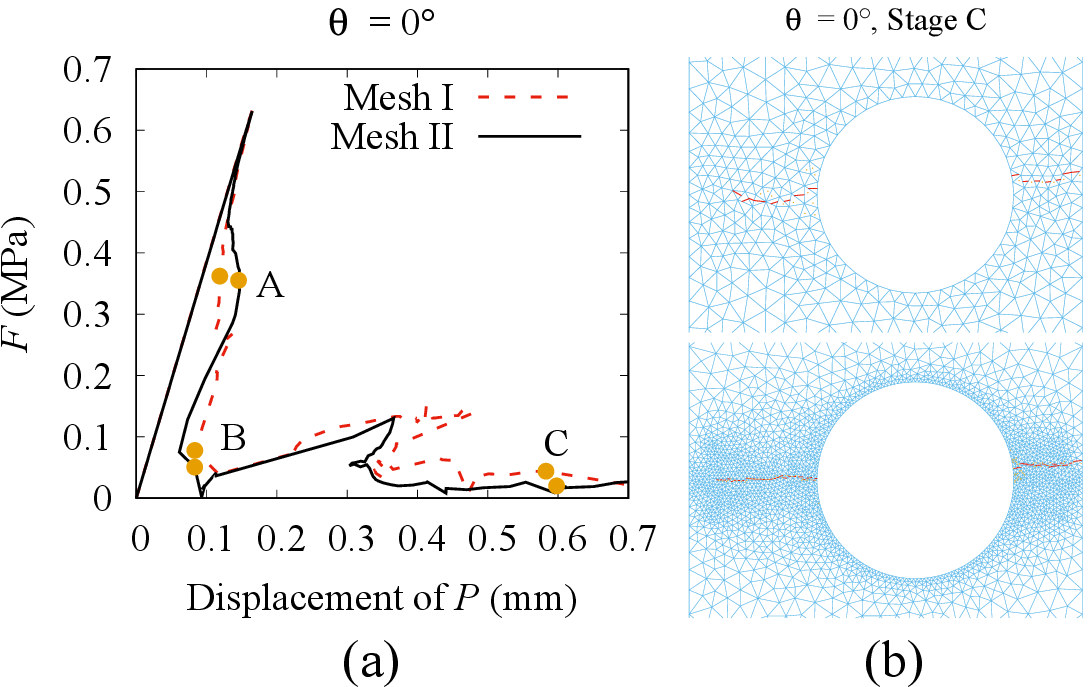}
	\caption{Perforated plate with a hole, case $\theta=0^\circ$: (a) force-displacement curves, (b) crack paths at stage C}
	\label{fig:Rech0degF}
\end{figure}

\begin{figure}[htbp]
	\centering
	\includegraphics[width=0.75\textwidth]{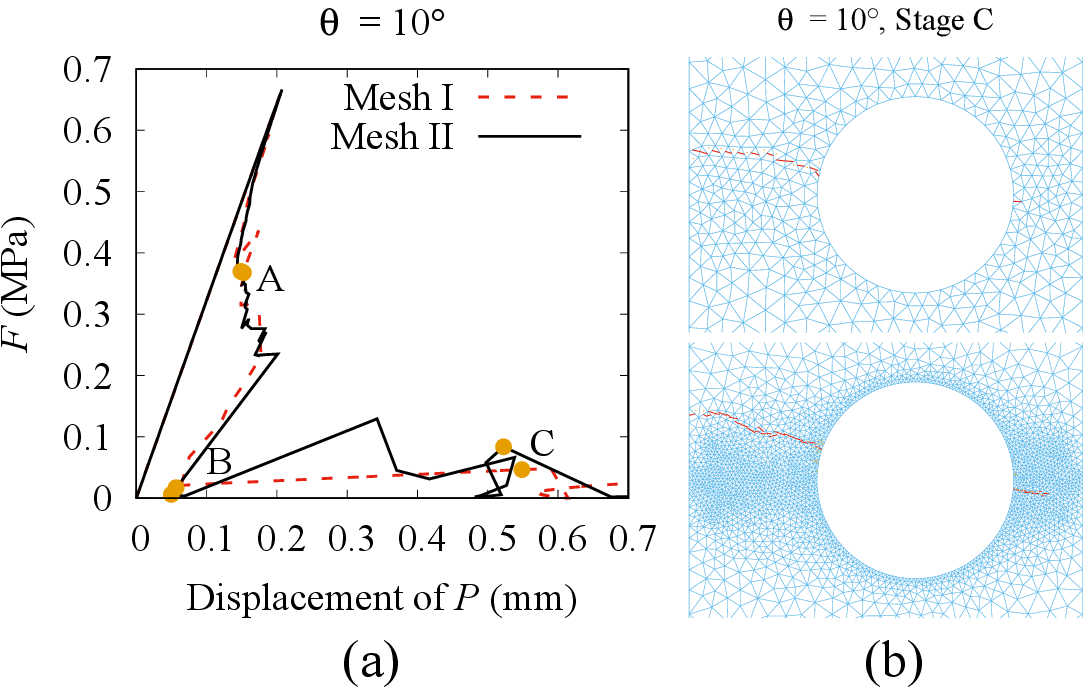}
	\caption{Perforated plate with a hole, case $\theta=10^\circ$: (a) force-displacement curves, (b) crack paths at stage C}
	\label{fig:Rech10degF}
\end{figure}

\begin{figure}[htbp]
	\centering
	\includegraphics[width=0.75\textwidth]{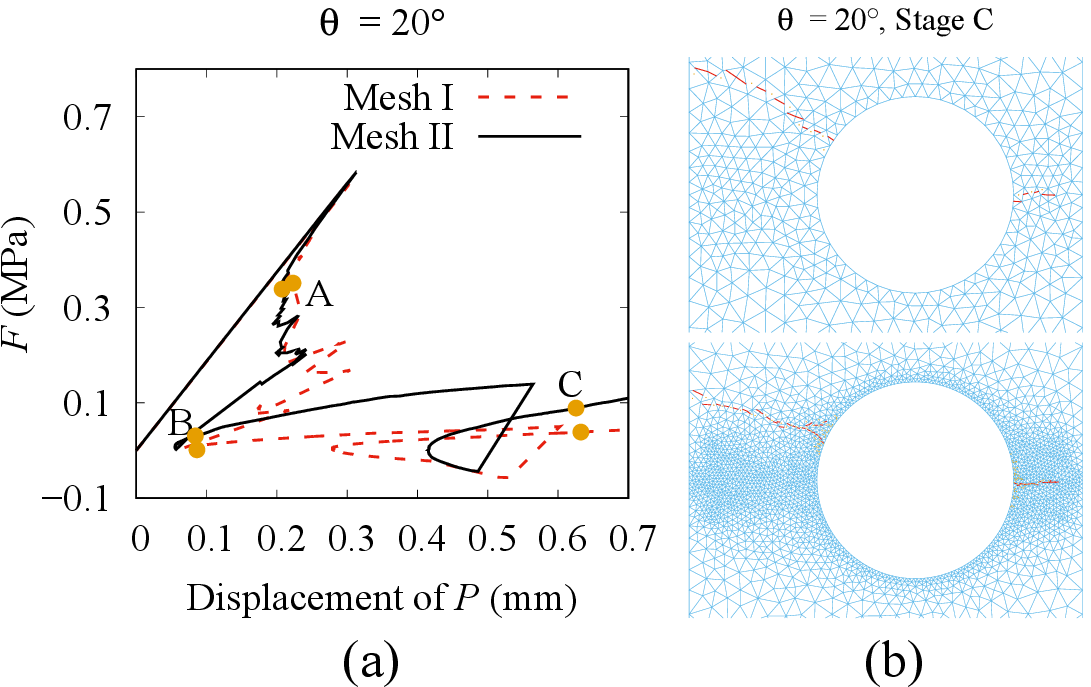}
	\caption{Perforated plate with a hole, case $\theta=20^\circ$: (a) force-displacement curves, (b) crack paths at stage C}
	\label{fig:Rech20degF}
\end{figure}

\begin{figure}[htbp]
	\centering
	\includegraphics[width=0.75\textwidth]{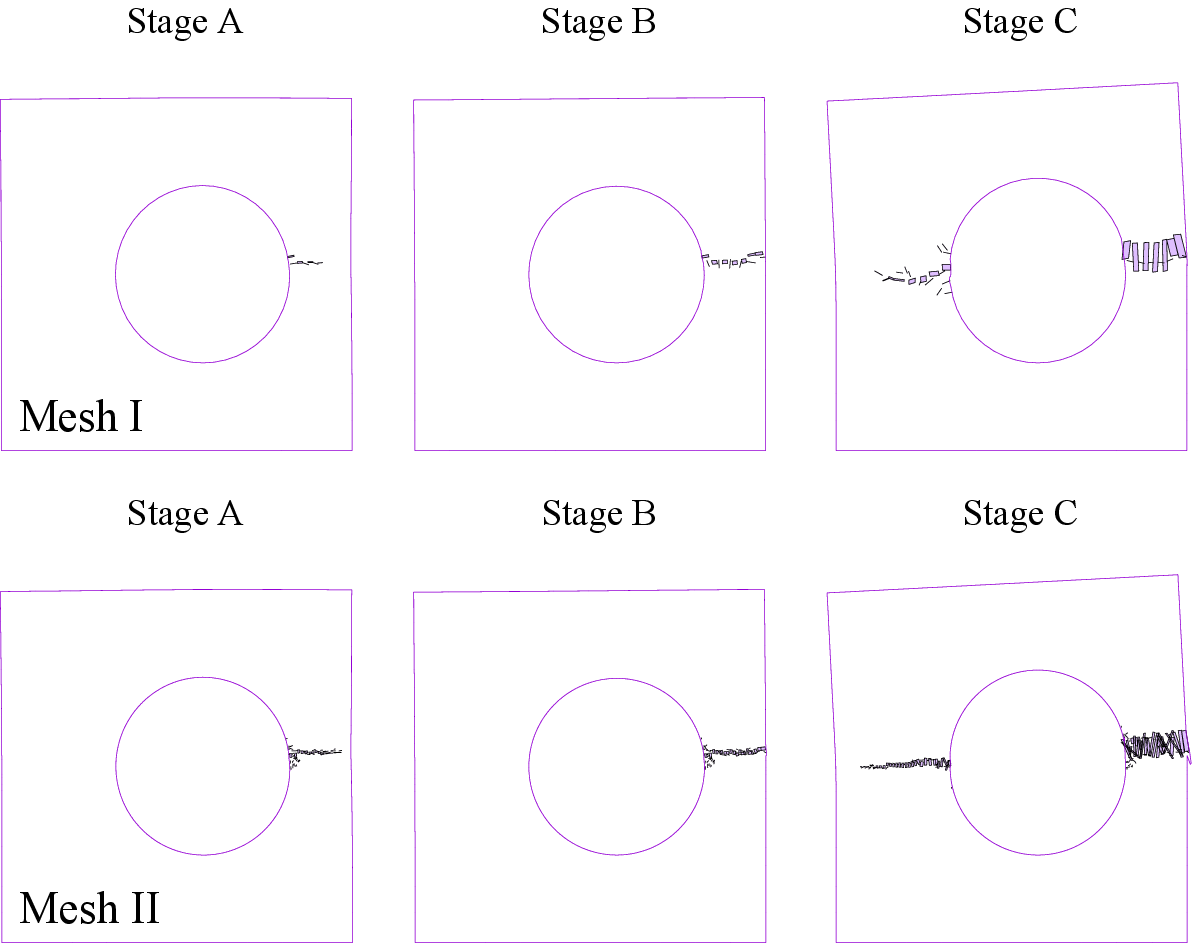}
	\caption{Perforated plate with a hole, case $\theta=0^\circ$: crack opening plots with deformed boundaries (scale = 1:100)}
	\label{fig:Rech0degCW}
\end{figure}

\begin{figure}[htbp]
	\centering
	\includegraphics[width=0.75\textwidth]{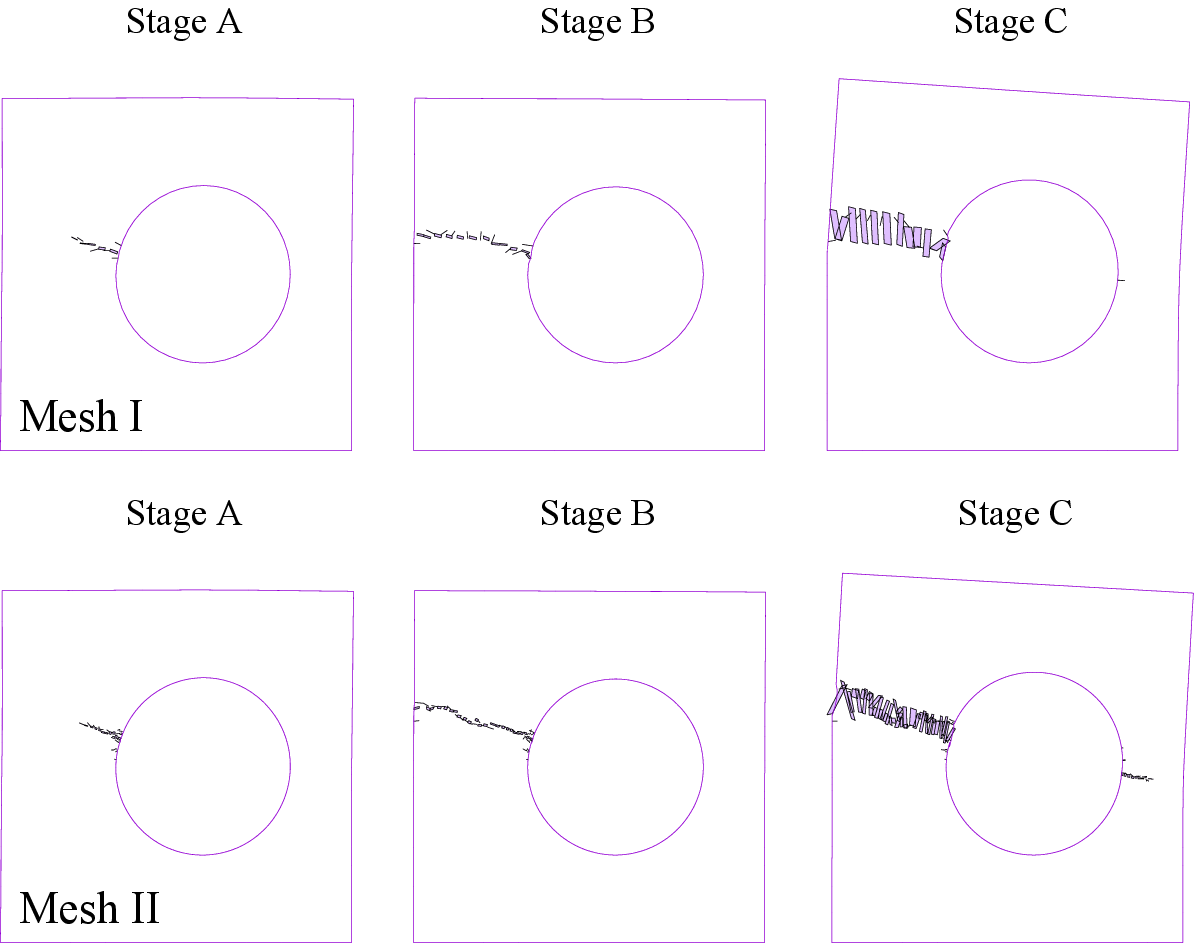}
	\caption{Perforated plate with a hole, case $\theta=10^\circ$: crack opening plots with deformed boundaries (scale = 1:100)}
	\label{fig:Rech10degCW}
\end{figure}

\begin{figure}[htbp]
	\centering
	\includegraphics[width=0.75\textwidth]{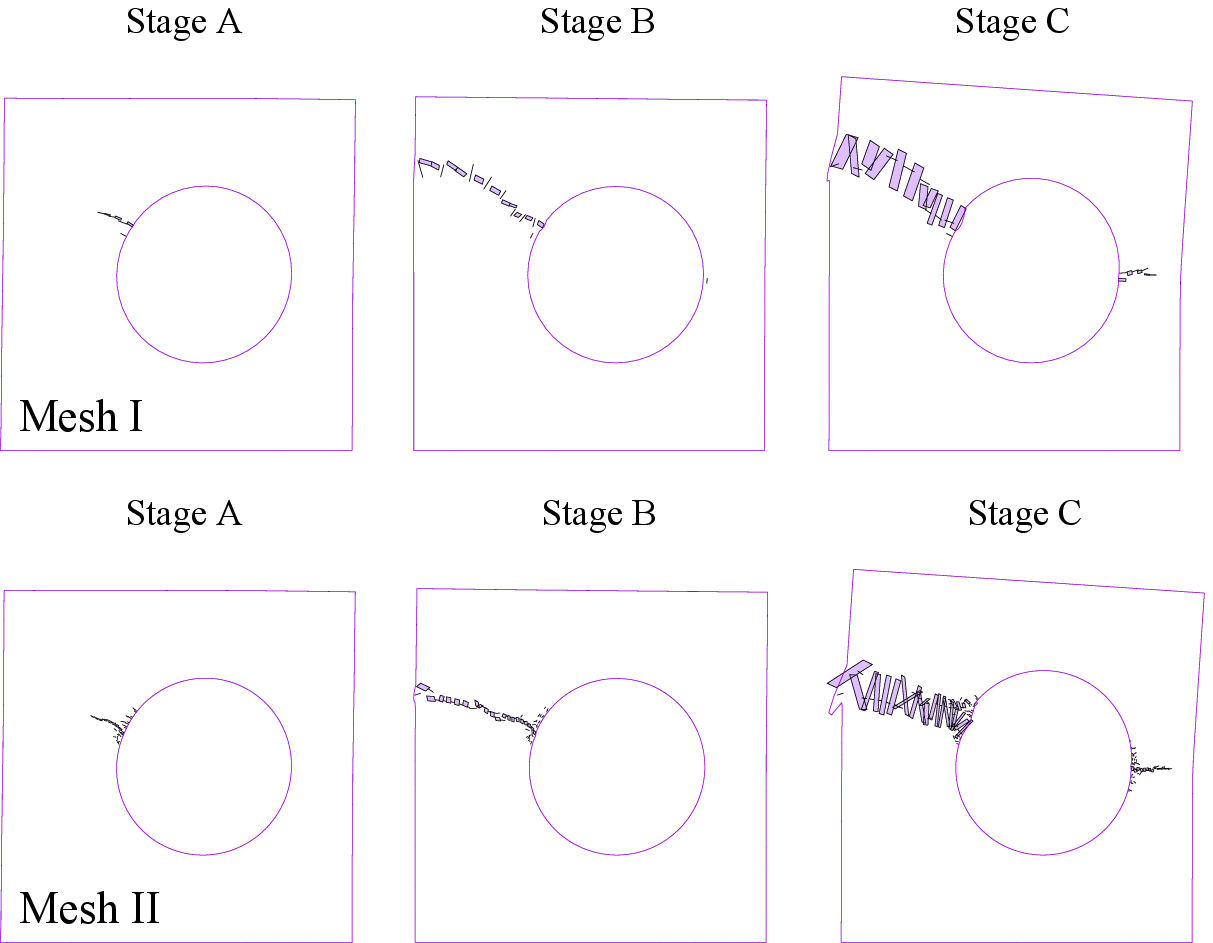}
	\caption{Perforated plate with a hole, case $\theta=20^\circ$: crack opening plots with deformed boundaries (scale = 1:100)}
	\label{fig:Rech20degCW}
\end{figure}

Three cases with inclined angles $\theta=0^\circ$, $\theta=10^\circ$, and $\theta=20^\circ$ of the distributed force are considered, leading to different force-displacement curves and crack paths, as shown in Figures~\ref{fig:Rech0degF} -~\ref{fig:Rech20degF}.  For the cases with $\theta=10^\circ$ and $\theta=20^\circ$, oscillations of the force-displacement curves are observed.  They are the consequence of the propagation of the cracks in relatively coarse meshes.  When the stress states at the center points of the elements are used to determine whether the element experiences cracking, coarse meshes may result in postponing local cracking \cite{Yiming:11,Yiming:14}.  Crack opening and force-displacement curves are shown in Figures~\ref{fig:Rech0degCW} -~\ref{fig:Rech20degCW}.  In case of $\theta=0^\circ$, the crack starts to propagate on the right side of the hole.  In case of $\theta=10^\circ$ and $\theta=20^\circ$, however, it starts to propagate on the left side.  In the proposed approach, the initiation and the propagation of the cracks are captured automatically.

\subsection{Specimen with multiple cracks}
\begin{figure}[htbp]
	\centering
	\includegraphics[width=0.95\textwidth]{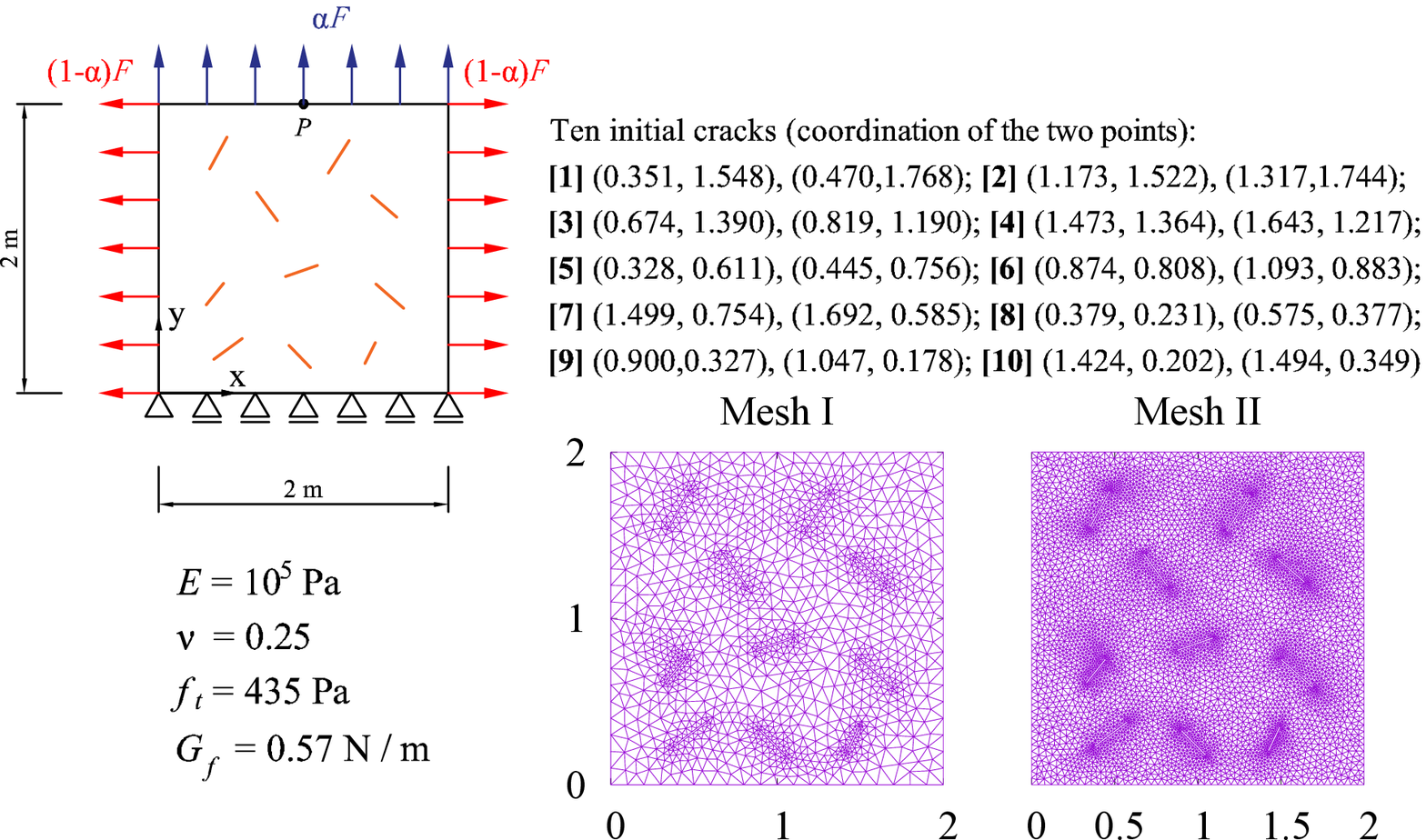}
	\caption{Specimen with multiple cracks: model, materials, and meshes}
	\label{fig:Multimodel}
\end{figure}

Quasi-brittle materials with multiple cracks, such as fractured rock, are common in geotechnical engineering practice.  The cracking process of such materials commonly exhibits a strong instability.  In this example, a plate with ten initial cracks is considered.  The model and the meshes are shown in Figure~\ref{fig:Multimodel}, where the width of all cracks is assumed to be 1~cm.  This example was investigated in \cite{Budyn2004,Zi_2004,RABCZUK201742} with XFEM and peridynamics.  In contrast to the literature, in this work not only uni-axial but also biaxial loads are considered.  As shown in Figure~\ref{fig:Multimodel}, a parameter $\alpha$ is used for controlling the differences between vertical and horizontal loads.  

\begin{figure}[htbp]
	\centering
	\includegraphics[width=0.9\textwidth]{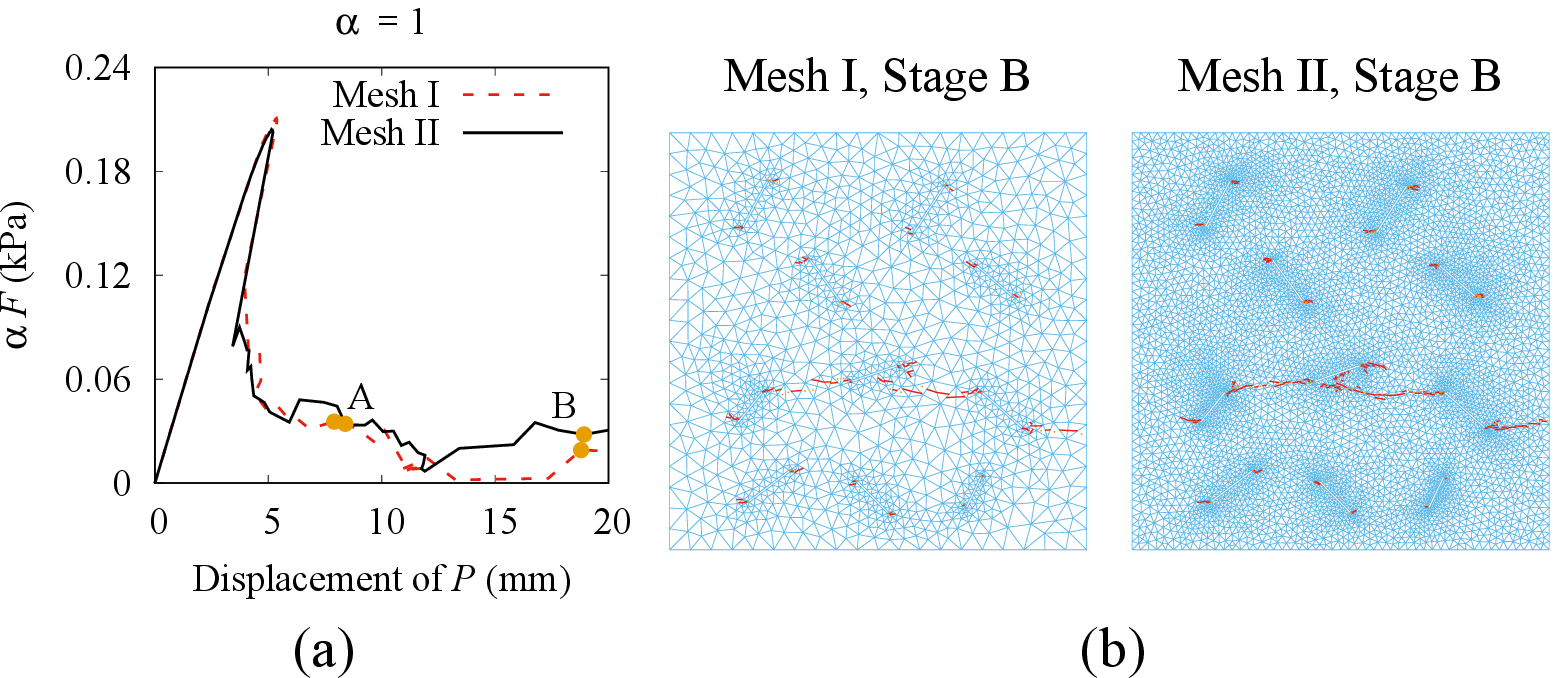}
	\caption{Specimen with multiple cracks with $\alpha=1$: (a) force-displacement curves, (b) crack paths at stage B}
	\label{fig:MultiFalpha1}
\end{figure}

\begin{figure}[htbp]
	\centering
	\includegraphics[width=0.9\textwidth]{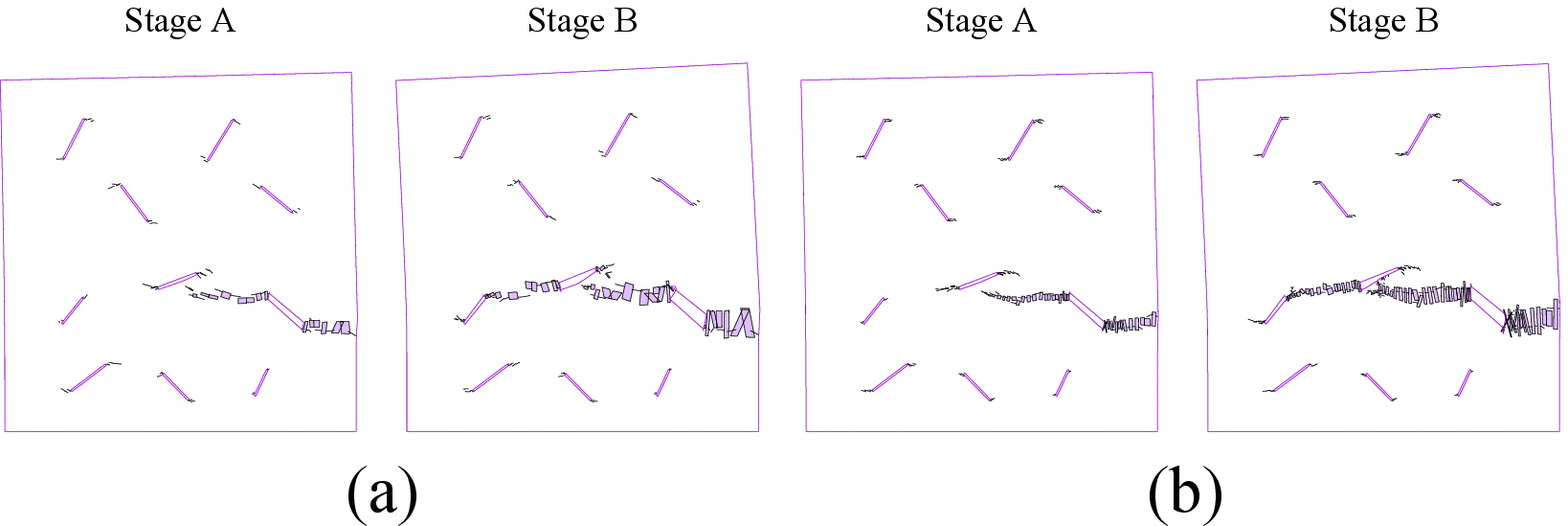}
	\caption{Specimen with multiple cracks with $\alpha=1$:  crack opening plots with deformed boundaries (scale = 1:4), (a) Mesh I, (b) Mesh II}
	\label{fig:MultiCWalpha1}
\end{figure}

For the case with $\alpha=1$, the specimen is loaded uni-axially.  The force-displacement curves and crack paths are shown in Figure~\ref{fig:MultiFalpha1}, and the crack opening and deformation plots are shown in Figure~\ref{fig:MultiCWalpha1}. These results are consistent with the ones in the literature.  Furthermore, all figures show similar results for different meshes.  

\begin{figure}[htbp]
	\centering
	\includegraphics[width=0.9\textwidth]{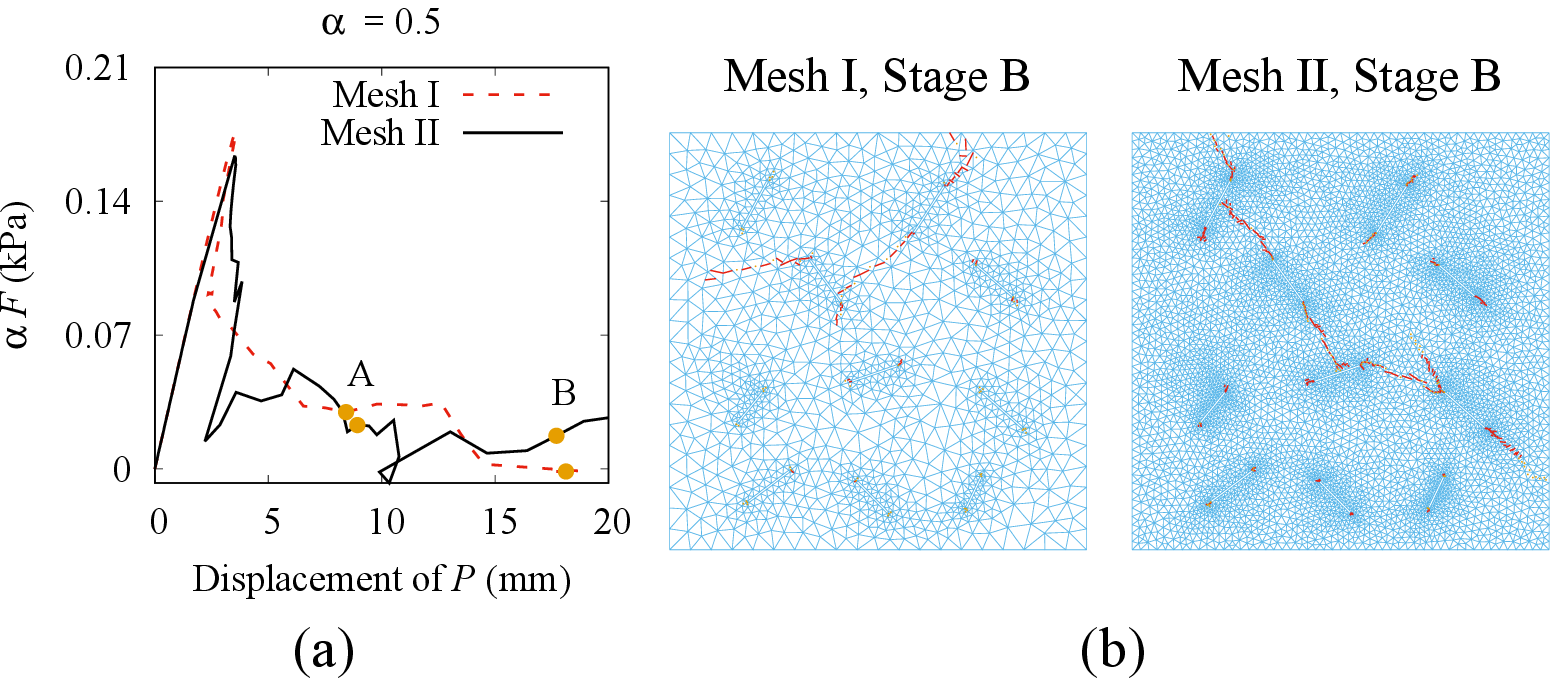}
	\caption{Specimen with multiple cracks with $\alpha=0.5$: (a) force-displacement curves, (b) crack paths at stage B}
	\label{fig:MultiFalpha05}
\end{figure}

\begin{figure}[htbp]
	\centering
	\includegraphics[width=0.9\textwidth]{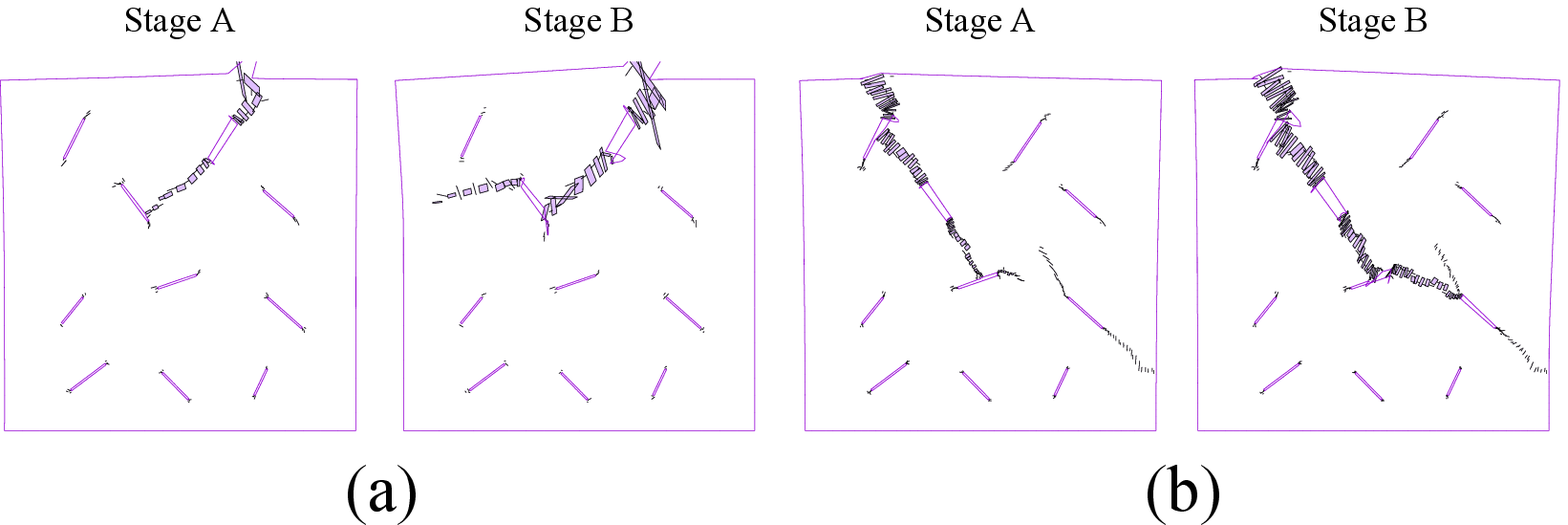}
	\caption{Specimen with multiple cracks with $\alpha=0.5$: crack opening plots with deformed boundaries (scale = 1:4), (a) Mesh I, (b) Mesh II}
	\label{fig:MultiCWalpha05}
\end{figure}

\begin{figure}[htbp]
	\centering
	\includegraphics[width=0.7\textwidth]{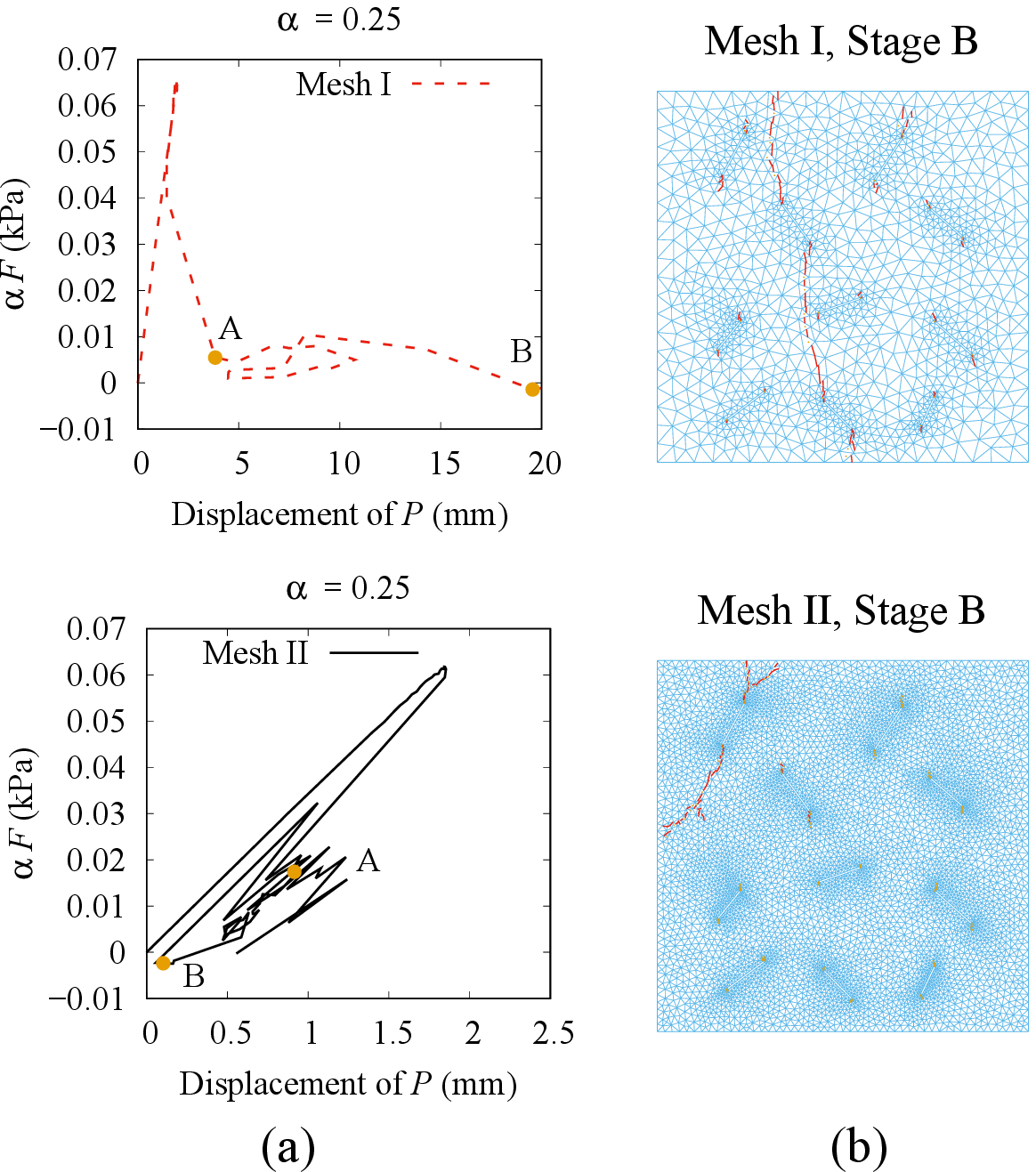}
	\caption{Specimen with multiple cracks with $\alpha=0.25$: (a) force-displacement curves, (b) crack paths at stage B}
	\label{fig:MultiFalpha025}
\end{figure}

\begin{figure}[htbp]
	\centering
	\includegraphics[width=0.9\textwidth]{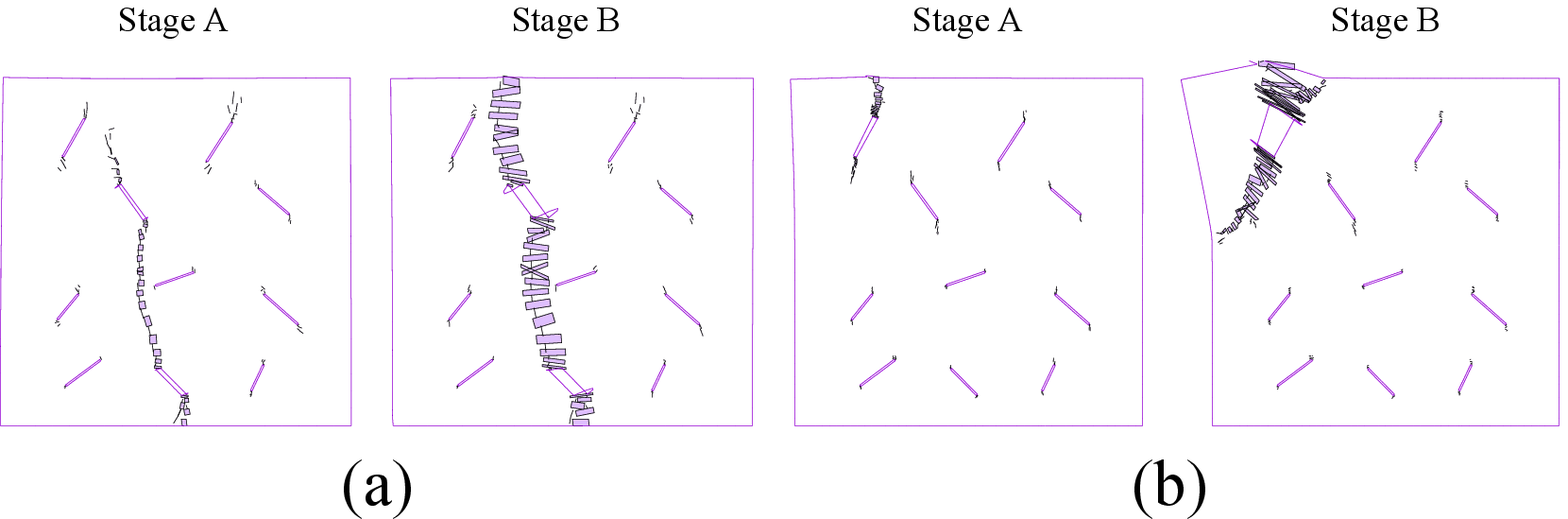}
	\caption{Specimen with multiple cracks with $\alpha=0.25$: crack opening plots with deformed boundaries (scale = 1:4), (a) Mesh I, (b) Mesh II}
	\label{fig:MultiCWalpha025}
\end{figure}

For the cases with $\alpha=0.5$ and $\alpha=0.25$, the specimen is loaded biaxially.  The force-displacement curves and crack paths are shown in Figures~\ref{fig:MultiFalpha05} and~\ref{fig:MultiFalpha025}, and crack opening plots with deformed boundaries are shown in Figures~\ref{fig:MultiCWalpha05} and~\ref{fig:MultiCWalpha025}.  Different crack paths and force-displacement curves are obtained from different meshes.  However, the peak values of the force-displacement curves from different meshes are very similar.  For such complex loading conditions, many elements, surrounding several crack tips experience stress concentrations with similar magnitudes.  Different meshes provide results with similar but different stress states.  This may lead to very different crack propagations.  This flaw of mesh dependence suggests further investigations.  Still, regarding the proposed direct dissipation-based arc-length approach, the mechanical response of all cases are captured well.

\section{Conclusions}
\label{sec:conc}
In this article, a direct dissipation-based arc-length approach was proposed in order to capture and trace the damage process of structures made of quasi-brittle materials in a stable and robust manner.  In contrast to the internal energy and work done by the external loads, the dissipated energy was used directly in the proposed approach.  It served as the monotonically increasing state variable.  Moreover, when combined with the Sherman-Morrison formula, the stiffness factor of the arc-length restraint was naturally obtained.  The approach was implemented in the framework of the CEM.  Its effectiveness and reliability was demonstrated by several benchmark tests, where the cracks were self-propagating and the force-displacement curves were continuously traced.

\section{Acknowledgement}
The authors gratefully acknowledge financial support by the National Natural Science Foundation of China (NSFC) (51809069) and by the Hebei Province Natural Science Fund E2019202441.